\documentclass[ceqn, prd, showkeys, prd, onecolumn, nofootinbib, amsmath, amsfonts, amssymb]{revtex4-1}

\usepackage{amssymb,amsmath,mathtools,xcolor,graphicx,xspace,colortbl,ragged2e,rotating} %
\usepackage{textcomp}
\usepackage{amsfonts}  
\usepackage{amsmath}  
\usepackage{amssymb}  
\usepackage{boxedminipage}  
\usepackage{graphics}  
\usepackage{graphicx}  
\usepackage{ragged2e}  
\usepackage{tabulary}  
\usepackage{wrapfig}  
\usepackage{xcolor}  
\usepackage{varioref}  
\graphicspath{{FRACTIONAL6_1_graphics/}{FRACTIONAL6_1_tcache/}{FRACTIONAL6_1_gcache/}}
\DeclareGraphicsExtensions{.pdf,.eps,.ps,.png,.jpg,.jpeg}

\begin{document}
\title{NEWTONIAN FRACTIONAL-DIMENSION GRAVITY AND THE EXTERNAL FIELD EFFECT}
\author{Gabriele U. Varieschi}
\email[E-mail me at: ]{Gabriele.Varieschi@lmu.edu}
\homepage[Visit: ]{http://gvarieschi.lmu.build}
\affiliation{Department of Physics, Loyola Marymount University, Los Angeles, CA 90045, USA}

\date{
\today
}

\begin{abstract}
We expand our analysis of Newtonian Fractional-Dimension Gravity (NFDG), an extension of the classical laws of Newtonian gravity to lower dimensional spaces, including those with fractional
(i.e., non-integer) dimension. We apply our model to four rotationally supported galaxies (NGC 5033, NGC 6674, NGC 5055, NGC 1090), in addition to other three galaxies (NGC 7814, NGC 6503, NGC 3741) which were analyzed in previous studies. NFDG is able to fit the rotation curves of all these galaxies without any dark matter component.

We also investigate the possible violation of the strong equivalence principle, in relation to the External Field Effect (EFE), i.e., the dependence of the internal motion of a self-gravitating system under freefall on an external gravitational field.  This effect is not present in Newtonian or Einstein gravity, but is predicted by some alternative theories of gravity. On the contrary, we show that NFDG does not imply the EFE, at least for values of the fractional dimension in the range $1 \leq D \leq 3$.

Using improved NFDG numerical computations, we analyze the rotation curves of the aforementioned galaxies and obtain perfect fits to the experimental data, by using a fractional-dimension function $D\left (R\right )$ which characterizes each individual galaxy. In the galactic sample studied here with NFDG methods, we do not detect any significant differences between galaxies that are supposed to show/not show the EFE according to other alternative theories. A larger sample of galaxies will be needed to fully determine the absence of any external field effect in NFDG.
\end{abstract}

\maketitle



\section{Introduction}
\label{sect:intro}

In this work, we continue the analysis started in previous publications (\cite{Varieschi:2020ioh,Varieschi:2020dnd,Varieschi:2020hvp,Varieschi:2021rzk}, hereafter papers I-IV, respectively) of Newtonian Fractional-Dimension Gravity (NFDG), an alternative gravitational model based on the assumption that Newtonian gravity can be extended to astrophysical objects characterized by a fractional dimension $D <3$, i.e., different from the standard value of the spatial dimension $D =3$.

The main goal of NFDG is to model galactic dynamics without any Dark Matter (DM) component, as it is done by some of the other existing alternative gravities (Modified Newtonian Dynamics - MOND \cite{Milgrom:1983ca,Milgrom:1983pn,Milgrom:1983zz}, Conformal Gravity - CG \cite{Mannheim:1988dj,Mannheim:1992tr,Mannheim:2005bfa}, Modified Gravity - MOG \cite{Moffat:2005si}, just to name a few). With regard to this objective, NFDG has achieved positive results thus far, but limited to just a few cases due to the recent introduction of this model. In paper I \cite{Varieschi:2020ioh}, we generalized the gravitational Gauss's law and used it to introduce the NFDG potential and field for spherically-symmetric structures; in paper II \cite{Varieschi:2020dnd}, we extended our analysis to axially-symmetric structures in order to analyze disk components of galaxies, and produced our first rotation curve fit for NGC 6503. In paper III \cite{Varieschi:2020hvp}, we included two more rotationally supported galaxies (NGC 7814 and NGC 3741), while in our paper IV \cite{Varieschi:2021rzk} we presented a relativistic version of our model (Relativistic Fractional-Dimension Gravity - RFDG).

In this paper, we will study four more galaxies with our methods  (NGC 5033, NGC 6674, NGC 5055, NGC 1090), using data from the Spitzer Photometry and Accurate Rotation Curves (SPARC) catalog \cite{Lelli:2016zqa}, and show that NFDG can produce perfect fits also in these cases.  In particular, these four galaxies were used in recent studies \cite{Chae:2020omu,Chae:2021dzt,Chae:2022rdr} based on Milgromian dynamics (i.e., MOND), to detect a possible violation of the Strong Equivalence Principle (SEP), by means of the so-called External Field Effect - EFE (see \cite{Mannheim:2021mhj} for a general overview, and references therein).

Newton's shell theorem for a spherical distribution of matter states: \textit{a body that is inside a spherical shell of matter experiences no net gravitational force from that shell}, and this statement is also valid in Einstein's General Relativity (GR). On the contrary, for most alternative theories of gravity \cite{Mannheim:2021mhj} the same statement does not hold anymore, and a test particle inside a spherical cavity cut out of a static, spherically-symmetric mass distribution will be subject to the EFE, due to the external mass distribution. However, we will show that NFDG verifies the Newtonian shell theorem, thus the EFE should not play any role within our model.

It should be noted that NFDG, as described in our papers I-IV, provides a method to perfectly fit the rotation curve $v_{circ}(R)$ of each galaxy by obtaining the corresponding variable dimension function $D(R)$, and then by using the NFDG gravitational potentials and fields to recompute the rotation curve, as described in Sect. \ref{sect:REVIEW} and in Appendix \ref{sectiongalacticappa}. This might raise issues about the falsifiability of our model, which seems to provide a mathematical connection between $v_{circ}(R)$ and $D(R)$ for each galaxy, without introducing more general principles to be used to derive the variable dimension function.

To address this point, in Appendix \ref{sectiongalacticdeep5055} we will present a simple method to determine $D(R)$ for each galaxy from first principles, based on the individual galactic mass distribution and on the concept of fractional mass dimension. Using this alternative procedure, we can obtain relatively good fits to the rotation data, although not as perfect as those obtained with our original procedure. We will still consider the variable dimension function $D(R)$ to be an individual characteristic of each galaxy, i.e., NFDG does not expect to find a universal $D(R)$ function, although galaxies of similar type might share similar dimension functions. Lastly, another possible way to falsify NFDG would be to obtain for a certain galaxy a dimension function which clearly does not make any physical sense, i.e., not in the standard range $1 \lesssim D(R) \lesssim 3$. This has not happened yet, for the sample of galaxies analyzed with our methods, but more cases will need to be studied to check this point.

The contents of the paper are organized as follows: in section \ref{sect:REVIEW} we will review the NFDG model and its connections with galactic dynamics and MOND. In section \ref{sectiongalacticefe}, we will discuss the external field effect and the differences between NFDG and some other alternative theories of gravity. Our new NFDG galactic rotation curves for the four galaxies mentioned above will be presented in section \ref{sectiongalactic}. Slight changes and improvements in our NFDG computations will be outlined in Appendix \ref{sectiongalacticappa}, while the other three galaxies from previous papers will be briefly revisited in Appendix \ref{sectiongalacticfive}. An extended analysis of NGC 5055 will be presented in Appendix \ref{sectiongalacticdeep5055}, to address particular issues such as the falsifiability of our model, the error band for the dimension function $D(R)$, and the sensitivity of NFDG results to a possible EFE detection. Final conclusions are reported in section \ref{sectiongalacticconcl}.

\section{NFDG and galactic dynamics}
\label{sect:REVIEW}

In this section we will summarize the fundamental ideas of NFDG, together with its applications to galactic dynamics. More details can be found in papers I-IV \cite{Varieschi:2020ioh,Varieschi:2020dnd,Varieschi:2020hvp,Varieschi:2021rzk} and references therein.

Newtonian Fractional-Dimension Gravity was based on a heuristic extension of Gauss's law for gravitation to a lower-dimensional space-time $D +1$, where $D \leq 3$ can be a non-integer space dimension. In paper I \cite{Varieschi:2020ioh}, we argued that the gravitational field due to a point mass $m$ placed at the origin in a $D$-dimensional space, should become:

\begin{equation}\mathbf{g}\left (r\right ) = -2\pi ^{1 -D/2}\Gamma \left (D/2\right )\frac{Gm}{l_{0}^{2}r^{D -1}} , \label{eq2.1}
\end{equation}where $G$ is the gravitational constant and $l_{0}$ is an appropriate scale length, which is required for dimensional correctness in fractional gravity models. In Eq. (\ref{eq2.1}) the radial coordinate $r$ is considered to be a dimensionless coordinate, obtained by rescaling the standard radial coordinate, i.e., $r \rightarrow r/l_{0}$ in all formulas in this paper.\protect\footnote{
	In our previous papers \cite{Varieschi:2020ioh,Varieschi:2020dnd,Varieschi:2020hvp,Varieschi:2021rzk}, we used the more cumbersome notation $w_{r} =r/l_{0}$ and similar notation for all the other rescaled coordinates, as well as denoting with $\widetilde{m}_{\left (D\right )}$ the rescaled point mass (in kilograms) in a $D$-dimensional space. In this paper we prefer to use a simplified notation, with $r$ and $m$ denoting these rescaled quantities. SI units will also be used throughout this paper, unless otherwise noted.
}

The fundamental result in Eq. (\ref{eq2.1}) above was based on the generalized expression $\int \nolimits_{S}d\Omega _{D} =2\pi ^{D/2}/\Gamma (D/2)$, for a hypersphere $S$ in a $D$-dimensional space used as a Gaussian surface, and on the related generalization of the integral for a spherically-symmetric function $f\left (r\right )$ over a $D$-dimensional space $W$: $\int \nolimits_{W}fd\mu _{H} =\frac{2\pi ^{D/2}}{\Gamma \left (D/2\right )}\int \nolimits_{0}^{\infty }f\left (r\right )r^{D -1}dr$, where $\mu _{H}$ denotes an appropriate Hausdorff measure over the space. 

It is easy to check that all the equations above reduce to standard results for the case $D =3$, and it is also possible to introduce a generalized gravitational potential as follows:
\begin{gather}\Phi (r) = -\frac{2\pi ^{1 -D/2}\Gamma (D/2)\ Gm}{\left (D -2\right )l_{0}r^{D -2}}\ ;\ D \neq 2 \label{eq2.2} \\
	\Phi \left (r\right ) =\frac{2\ Gm}{l_{0}}\ln r\ ;\ D =2 , \nonumber \end{gather}
with the connection between gravitational field and potential given as: $\mathbf{g}\left (r\right ) = - \nabla \Phi \left (r\right )/l_{0}$, where $ \nabla $ denotes the standard gradient operator acting over the dimensionless coordinate $r$, so that the scale length $l_{0}$ is needed to obtain the correct physical dimensions.\protect\footnote{
	Comparing with our papers I-IV, we also simplified the notation for the potential $\Phi \left (r\right )$ and the gradient $ \nabla $, which used to be denoted respectively by $\widetilde{\phi }\left (w\right )$ and $ \nabla _{D}$ in previous work.
}

More generally, we can consider a fractional-dimension Poisson equation:

\begin{equation} \nabla _{D}^{2}\Phi  =\frac{4\pi G}{l_{0}}\rho  , \label{eq2.3}
\end{equation}where $ \nabla _{D}^{2}$ is a generalized $D$-dimensional Laplacian (see Ref. \cite{Varieschi:2020ioh}, or the more detailed discussion in Ref. \cite{Varieschi:2021rzk}), and $\rho $ represents here the rescaled mass density in
$\mbox{kg}\thinspace $.\protect\footnote{
	Again, due to our change of notation, this current Eq. (\ref{eq2.3}) looks different from Eq. (22) in Ref. \cite{Varieschi:2020ioh}, but the two Poisson equations are in fact equivalent.
}
The solution to this Poisson equation, for a point-like mass $m$ placed at the source position $\mathbf{x}^{ \prime }$, yields the gravitational field $\Phi \left (\mathbf{x}\right )$ at the field position $\mathbf{x}$ as:

\begin{equation}\Phi \left (\mathbf{x}\right ) =m\mathcal{G}\left (\mathbf{x} -\mathbf{x}^{ \prime }\right ) , \label{eq2.4}
\end{equation}factoring out the dependence on the mass $m$. The Green's function $\mathcal{G}$ follows from Eq. (\ref{eq2.2}):
\begin{gather}\mathcal{G}\left (\mathbf{x} -\mathbf{x}^{ \prime }\right ) = -\frac{2\pi ^{1 -D/2}\Gamma (D/2)G\ }{\left (D -2\right )l_{0}\left \vert \mathbf{x} -\mathbf{x}^{ \prime }\right \vert ^{D -2}}\ ;\ D \neq 2 \label{eq2.5} \\
	\mathcal{G}\left (\mathbf{x} -\mathbf{x}^{ \prime }\right ) =\frac{2G\ }{l_{0}}\ln \left \vert \mathbf{x} -\mathbf{x}^{ \prime }\right \vert \ ;\ D =2. \nonumber \end{gather}

Considering instead an extended source mass distribution, such as a galaxy, globular cluster, or other astrophysical objects where these methods might be applicable, the full gravitational potential is obtained by replacing the point mass $m$ with an infinitesimal mass $dm =\rho \left (\mathbf{x}^{ \prime }\right )d\mathcal{V}$ and then by integrating over the total volume $\mathcal{V}$:
\begin{equation}\Phi \left (\mathbf{x}\right ) =\int \nolimits_{\mathcal{V}}d^{3}\mathbf{x}^{ \prime }\rho \left (\mathbf{x}^{ \prime }\right )\mathcal{G}\left (\mathbf{x} -\mathbf{x}^{ \prime }\right ) . \label{eq2.6}
\end{equation}

A further complication of NFDG \cite{Varieschi:2020ioh,Varieschi:2020dnd,Varieschi:2020hvp,Varieschi:2021rzk}, and of more general fractional spacetime theories \cite{Calcagni:2021mmj}, is that the volume integral over a $D$-dimensional space must be performed by including an appropriate weight $v\left (\mathbf{x}^{ \prime }\right )$ inside the metric of the space, i.e.:
\begin{equation}\Phi _{NFDG}\left (\mathbf{x}\right ) =\int \nolimits_{\mathcal{V}}d^{3}\mathbf{x}^{ \prime }v\left (\mathbf{x}^{ \prime }\right )\rho \left (\mathbf{x}^{ \prime }\right )\mathcal{G}\left (\mathbf{x} -\mathbf{x}^{ \prime }\right ) , \label{eq2.7}
\end{equation}where the form of the weight depends on the type of fractional theory and also on the type of coordinates (spherical, cylindrical, etc.).

If the astrophysical object being studied contains different mass components (and with different overall symmetries), the NFDG potential will be simply the sum of all the different contributions, each computed with an appropriate integral of the general form (\ref{eq2.7}). The gravitational potential in Eqs. (\ref{eq2.5}) and (\ref{eq2.7}) was originally introduced for a fixed value of the fractional dimension $D$, but in papers I-IV it was argued that it could be applicable also to the case of a variable dimension $D\left (\mathbf{x}\right )$, assuming a slow change of the dimension $D$ with the field point coordinates.

In Appendix \ref{sectiongalacticappa}, we will include full details of the NFDG\ procedure used to compute the integrals in Eq. (\ref{eq2.7}) for the spherically/cylindrically-symmetric structures in the galaxies considered in this paper, which typically have a spherical bulge mass distribution and cylindrical stellar disk and gas distributions. In all these cases, the resulting gravitational potential in the galactic disk plane $\Phi _{NFDG}\left (R\right )$ will be a function of the (dimensionless) radial coordinate $R$ and the NFDG gravitational field will be computed by simple differentiation:\protect\footnote{
	Although the variable dimension $D\left (R\right )$ will be considered a function of the radial coordinate $R$, we will assume $\frac{dD}{dR} \approx 0$, in view of the assumption of a slowly varying fractional dimension over galactic radial distances.
}

\begin{equation}\mathbf{g}_{NFDG}\left (R\right ) = -\frac{1}{l_{0}}\frac{d\Phi _{NFDG}\left (R\right )}{dR}\widehat{\mathbf{R}} , \label{eq2.8}
\end{equation}with the scale length $l_{0}$ included for dimensional correctness, since $R$ is also considered as a dimensionless coordinate.

The circular velocity, for stars rotating in the main galactic plane, is simply obtained from Eq. (\ref{eq2.8}) as:

\begin{equation}v_{circ}\left (R\right ) =\sqrt{l_{0}R\left \vert \mathbf{g}_{NFDG}\left (R\right )\right \vert } \label{eq2.9}
\end{equation}and the galactic rotation curves in Sect. \ref{sectiongalactic} will be obtained by plotting these circular velocities as a function of the radial distance from the galactic center.

The scale length $l_{0}$ is still needed in the previous equations to ensure dimensional correctness of all expressions for $D \neq 3$, even using the simplified notation of this paper. However, as it will be shown in Appendix \ref{sectiongalacticappa}, the final results of our galactic fits are totally independent of the $l_{0}$ value, so that the only ``free parameter'' of our model is the variable dimension function $D\left (R\right )$, which becomes a key signature of each galaxy being analyzed.

In papers I-IV, the scale length $l_{0}$ was related to the MOND\  acceleration constant
$a_{0}$ as:
\begin{equation}a_{0} \approx \frac{GM}{l_{0}^{2}} , \label{eq2.10}
\end{equation}where $M$ is the total mass (or a reference mass) of the system being studied. The currently accepted value of the MOND\ constant (also denoted by $g_{\dag }$ \cite{McGaugh:2016leg,Lelli:2017vgz}) is:

\begin{equation}a_{0} \equiv g_{\dag } =\ 1.20 \times 10^{ -10}\ \mbox{}\ \mbox{m}\thinspace \mbox{s}^{ -2} , \label{eq2.11}
\end{equation}
and it represents the acceleration scale below which MOND corrections are needed.

We recall that MOND \cite{Milgrom:1983ca,Milgrom:1983pn,Milgrom:1983zz} introduced modifications of Newtonian dynamics in terms of modified inertia (MI), or modified gravity (MG)
\cite{Bekenstein:1984tv}. In particular, Milgrom's law
$m\mu (a/a_{0})\mathbf{a} =\mathbf{F}$ represents MI, as the mass $m$ is replaced by $m\mu \left (a/a_{0}\right )$, while its alternative version $\mu (g/a_{0})\mathbf{g} =\mathbf{g}_{N}$ represents MG, as the observed gravitational field $\mathbf{g}$ might differ from the Newtonian one $\mathbf{g}_{N}$. There is some evidence \cite{Petersen:2020vks,Petersen:2019mfm,Milgrom:2012rk} favoring MG over MI, but the two formulations are practically equivalent. However, conceptually,
MI is a modification of Newton's laws of motion, while
MG is a modification of Newton's law of universal gravitation. As already remarked in our papers I-IV, and following Eqs. (\ref{eq2.1})-(\ref{eq2.2}), NFDG will also be considered a modification of gravity and not of inertia. In NFDG we assume that test objects, such as stars in galaxies, will move in a
(classical)
$3 +1$
space-time and will obey standard laws of dynamics.

In paper I, the main consequences of the MOND\ theory were obtained in NFDG by considering the Deep-MOND regime as equivalent to a space dimension $D \approx 2$. In particular, the flat rotation velocity
$V_{f} \approx \sqrt[{4}]{GMa_{0}}$ exhibited by galactic rotation curves, the ``baryonic'' Tully-Fisher relation-BTFR:
$M_{bar} \sim V_{f}^{4}$, and other basic MOND predictions were recovered in NFDG by assuming $D \approx 2$  \cite{Varieschi:2020ioh}. The other key element of the MOND\ paradigm is the interpolating function\textit{} ($\mu (x) \equiv \mu (a/a_{0})\text{}$
for the MI\  case, or
$\mu \text{}(x) \equiv \mu (g/a_{0})$ for the MG case), with the two asymptotic behaviors postulated by MOND: $\mu \left (x\right ) \rightarrow 1\text{  for }x \gg 1\text{  (Newtonian regime)}$ and $\mu \left (x\right ) \rightarrow x$ $\text{for }x \ll 1\text{  (Deep-MOND regime)}$.

The connection between these two limiting regimes is usually a non-linear interpolating function $\mu \left (x\right )$ (or the inverse function $\nu \left (y\right )$, see papers I-II for details), with the favorite choice in the literature \cite{McGaugh:2016leg,Lelli:2017vgz} expressed as
$\nu (y) =\left [1 -\exp \left ( -y^{1/2}\right )\right ]^{ -1}$. This function yields the so-called Radial Acceleration Relation - RAR:

\begin{equation}g_{obs} =\frac{g_{bar}}{1 -e^{ -\sqrt{g_{bar}/g_{\dag }}}} , \label{eq2.12}
\end{equation}
connecting the
\textit{observed} radial acceleration
$g_{obs}$,
traced by rotation curves, with the \textit{baryonic }radial acceleration
$g_{bar}$,
predicted by the observed distribution of matter in galaxies, and with
$g_{\dag }$ given in Eq. (\ref{eq2.11})
above. The RAR \cite{McGaugh:2016leg,Lelli:2017vgz}  was originally introduced by analyzing data from 175 galaxies in the SPARC database \cite{Lelli:2016zqa} and was confirmed in more recent work  \cite{Lelli:2017vgz,Li:2018tdo} by adding
other early-type-galaxies (elliptical and lenticular) and dwarf spheroidal galaxies to the RAR analysis.

An improved version of the MOND-RAR interpolating function was recently introduced \cite{Chae:2020omu,Chae:2021dzt}, in view of a possible detection of the EFE. In the next section, we will review the EFE in connection with MOND\ and NFDG models, and discuss possible implications for galactic dynamics.

\section{NFDG and the External Field Effect}
\label{sectiongalacticefe}

The strong equivalence principle is a key feature of general relativity and can be used to test GR against possible alternative theories of gravity. The SEP\ implies that the internal motion of a self-gravitating system under freefall in an external gravitational field should not depend on the external field strength, except for possible tidal effects when the system being considered is not sufficiently small. MOND non-relativistic theory by Bekenstein and Milgrom \cite{Bekenstein:1984tv}, or the recent relativistic version by Skordis and Zlosnik \cite{Skordis:2020eui}, both imply violation of the SEP. In particular, they also violate local positional invariance for gravitational experiments, thus differentiating the SEP\ from the well tested Einstein equivalence principle \cite{Will:2014kxa}.

In view of this possible SEP violation of MOND, Chae et al. \cite{Chae:2020omu,Chae:2021dzt} have recently tested and reported an experimental detection of the EFE, within the paradigm of MOND dynamics, by observing this effect\  in individual ``golden'' galaxies (NGC 5033, NGC 5055) subject to strong external fields, and checking these results against exceptionally isolated galaxies (NGC 6674, NGC 1090), which represent ``control'' galaxies, for null detection of the EFE. In the following, we will also refer to these two classes of galaxies as ``EFE'' and ``non-EFE'' respectively.

The EFE was also statistically detected from a blind test of 153 galaxies of the SPARC catalog \cite{Chae:2020omu} and this analysis was improved in a second work on the subject \cite{Chae:2021dzt}, with 162 galaxies. The main point of the first analysis \cite{Chae:2020omu} was based on the general form of the MOND-RAR relation:

\begin{equation}g_{obs} =\nu _{0}\left (\frac{g_{bar}}{g_{\dag }}\right )g_{bar} , \label{eq3.1}
\end{equation}
of which previous Eq. (\ref{eq2.12}) is a special case, and $\nu _{0}\left (g_{bar}/g_{\dag }\right )$ now denotes a more general interpolating function. In general, Eq. (\ref{eq3.1}) is strictly valid only for isolated systems, when the EFE is negligible, and should be amended when the external field is not negligible anymore.

In particular, in Ref. \cite{Chae:2020omu} a \textit{Simple Interpolating Function} (IF) was used in the MOND-EFE case:

\begin{align}g_{MOND}\left (R\right ) =\nu _{e}\left (\frac{g_{bar}}{g_{\dag }}\right )g_{bar}\left (R\right ) \label{eq3.2} \\
	\nu _{e}\left (z\right ) =\frac{1}{2} -\frac{A_{e}}{z} +\sqrt{\left (\frac{1}{2} -\frac{A_{e}}{z}\right )^{2} +\frac{B_{e}}{z}} , \nonumber \end{align}
where $z \equiv g_{bar}/g_{\dag }$, $A_{e} \equiv e\left (1 +e/2\right )/\left (1 +e\right )$, $B_{e} \equiv \left (1 +e\right )$, and $e \equiv g_{ext}/g_{\dag }$, representing the external field strength compared with the MOND acceleration scale from Eq. (\ref{eq2.11}). For $e =0$ (non-EFE case) the Simple IF $\nu _{0}\left (z\right ) =1/2 +\sqrt{1/4 +1/z}$ is recovered. The above formulas were then used by Chae et al. \cite{Chae:2020omu} in their statistical analysis to fit individual EFE and non-EFE galaxies, as well as for a set of 153 SPARC\ galaxies.

Several free parameters were used in this analysis, including the galactic distance and disk inclination, the mass-to-light ratios for the galactic components (bulge, disk, gas), but with the external field parameter $e \equiv g_{ext}/g_{\dag }$ being the most important for the EFE detection. This parameter was also compared with typical values of the environmental gravitational parameter $e_{env} \equiv g_{env}/g_{\dag }$ ($e_{env} \approx 0.1$ for galaxies in dense environments, EFE galaxies; $e_{env} \approx 0.01$ for isolated galaxies, non-EFE).

The outcomes of this analysis \cite{Chae:2020omu} were that the EFE\  was detected in individual galaxies (NGC 5033 and NGC 5055) subject to strong external field, while not detected in control galaxies (NGC 1090, NGC 6674) residing in regions with weak external fields. The EFE was also statistically detected in a blind test with 153 SPARC\ galaxies and as a downward deviation from the empirical RAR \cite{Chae:2020omu}, while similar results were obtained in Ref. \cite{Chae:2021dzt} by using a more refined model of the Newtonian external field at each galactic position. These studies show that, at least within the MOND paradigm, there could be a breakdown of the SEP and a detection of the EFE, although the issue is still open \cite{Paranjape:2021eng}.

An alternative way of looking into the EFE for alternative theories of gravity was provided by Mannheim and Moffat \cite{Mannheim:2021mhj}, by considering the gravitational potentials and fields in the different theories and models. In Newtonian gravity, there is no force on a test particle located inside a spherical cavity cut out of a static, spherically symmetric mass distribution $\rho \left (r^{ \prime }\right )$, and the same applies to Einstein gravity. Therefore, there is no EFE acting on such test particle, unless a non-Newtonian potential is chosen, or for a geometry other than Ricci flat spacetime in relativistic theories.

The above statement is easily proved \cite{Mannheim:2021mhj}, for a Newtonian potential of the form:

\begin{equation}\Phi \left (r\right ) = -G\int d^{3}\mathbf{x}^{ \prime }\frac{\rho \left (r^{ \prime }\right )}{\left \vert \mathbf{x} -\mathbf{x}^{ \prime }\right \vert } , \label{eq3.3}
\end{equation}
from which, performing the angular integrations in spherical coordinates, one obtains:

\begin{equation}\Phi \left (r\right ) = -\frac{2\pi G}{r}\int dr^{ \prime }r^{ \prime }\rho \left (r^{ \prime }\right )\left [\left \vert r +r^{ \prime }\right \vert  -\left \vert r -r^{ \prime }\right \vert \right ] . \label{eq3.4}
\end{equation}

If the spherical mass distribution is non-zero only within a spherical shell of range $R_{1} <r^{ \prime } <R_{2}$, in the inner cavity ($r^{ \prime } <R_{1}$) and in the outer region ($r^{ \prime } >R_{2}$) we have, respectively:

\begin{align}\Phi _{in}\left (r\right ) = -4\pi G\int \nolimits_{R_{1}}^{R_{2}}dr^{ \prime }r^{ \prime }\rho \left (r^{ \prime }\right ) \label{eq3.5} \\
	\Phi _{out}\left (r\right ) = -\frac{4\pi G}{r}\int \nolimits_{R_{1}}^{R_{2}}dr^{ \prime }r^{ \prime 2}\rho \left (r^{ \prime }\right ) , \nonumber \end{align}showing that inside the cavity the gravitational potential is constant and, therefore, the field is zero (no EFE), while the standard $1/r$ behavior is recovered outside the spherical shell.

The first part of this statement (no EFE inside the cavity) remains true even if the spherically-symmetric mass distribution extends all the way to infinity ($R_{2} \rightarrow \infty $). For more realistic astrophysical and cosmological structures, which do not possess perfect spherical symmetry, tidal force effects are possible, typically falling off as $1/r^{3}$, and will be neglected in the EFE\ analysis.

From these initial considerations, Mannheim and Moffat \cite{Mannheim:2021mhj} then analyze a more general gravitational potential behaving as $\vert \mathbf{x} -\mathbf{x}^{ \prime }\vert ^{2\alpha  -1}$, with $\alpha  =0$ representing the previous Newtonian case, and easily show that for this type of potential the EFE\ is present inside the spherical cavity and the outer potential does not follow the $1/r$ dependence, except for the Newtonian ($\alpha  =0$) case. With additional considerations \cite{Mannheim:2021mhj}, based on the standard second-order Poisson equation and the equivalent Einstein equations, alternative theories of gravity such as MOND \cite{Milgrom:1983ca,Milgrom:1983pn,Milgrom:1983zz}, Conformal Gravity \cite{Mannheim:1988dj,Mannheim:1992tr,Mannheim:2005bfa}, and Modified Gravity \cite{Moffat:2005si}, are easily shown to imply an explicit EFE.

However, the situation in NFDG is different with regard to the external field effect. The NFDG potential is based on the Euler kernel, $1/\left \vert \mathbf{x} -\mathbf{x}^{ \prime }\right \vert ^{D -2}$ (see Appendix \ref{sectiongalacticappa}), which is similar to the potential $\vert \mathbf{x} -\mathbf{x}^{ \prime }\vert ^{2\alpha  -1}$ considered above and in Ref. \cite{Mannheim:2021mhj}, but the NFDG potential is obtained using the integrals in Eq. (\ref{eq8.1}), which include the NFDG\ weight $v\left (\mathbf{x}^{ \prime }\right )$ and, therefore, the EFE considerations above need to be revised.

We assume a spherically-symmetric mass distribution $\rho \left (r^{ \prime }\right )$ and align the field position vector $\mathbf{x}$ in the direction of the $z^{ \prime }$ axis, so that the angle between the field vector $\mathbf{x}$ and the source vector $\mathbf{x}^{ \prime }$ will simply be $\theta ^{ \prime }$, and thus $\vert \mathbf{x} -\mathbf{x}^{ \prime }\vert  =\sqrt{r +r^{ \prime } -2rr^{ \prime }\cos \theta ^{ \prime }}$. Following the details in Appendix \ref{sectiongalacticappa}, we then combine the general NFDG potential (for $D \neq 2$) in the first line of Eq. (\ref{eq8.1}), the weight in spherical coordinates from Eq. (\ref{eq8.6}), together with $d^{3}\mathbf{x}^{ \prime } =r^{ \prime 2}\sin \theta ^{ \prime }dr^{ \prime }d\theta ^{ \prime }d\phi ^{ \prime }$, $D =\alpha _{1} +\alpha _{2} +\alpha _{3}$, and obtain:

\begin{equation}\Phi _{NFDG}\left (r\right ) = -\frac{4\pi G}{\left (D -2\right )l_{0}}\int dr^{ \prime }r^{ \prime D -1}\rho \left (r^{ \prime }\right )\left (r^{2} +r^{ \prime 2}\right )^{1 -\frac{D}{2}}\,_{3}F_{2}\left (\begin{array}{ccc}\left (D -2\right )/4 , & D/4 , & \alpha _{3}/3 \\
		1/2 , & D/2 & \,\end{array} ;\frac{4r^{2}r^{ \prime 2}}{\left (r^{2} +r^{ \prime 2}\right )^{2}}\right ) , \label{eq3.6}
\end{equation}where $\,_{p}F_{q}\left (\begin{array}{c}a_{1} , . . . ,a_{p} \\
	b_{1} , . . . ,b_{q}\end{array} ;z\right ) \equiv \sum \limits _{k =0}^{\infty }\frac{\left (a_{1}\right )_{k} . . .\left (a_{p}\right )_{k}}{\left (b_{1}\right )_{k} . . .\left (b_{q}\right )_{k}}\frac{z^{k}}{k !}$ is the generalized hypergeometric series \cite{NIST:DLMF}.\protect\footnote{
	The Pochhammer's symbol
	$\left (a\right )_{k}$
	is defined as
	$\left (a\right )_{0} =1$,
	$\left (a\right )_{k} =a\left (a +1\right )\left (a +2\right ) . . .\left (a +k -1\right ) =\Gamma \left (a +k\right )/\Gamma \left (a\right )$.
} For $D =3$ ($\alpha _{1} =\alpha _{2} =\alpha _{3} =1$), the above expression effectively reduces to the one in Eq. (\ref{eq3.4}).\protect\footnote{
	The expression in Eq. (\ref{eq3.6}) represents another possible way to compute the NFDG potential/field for spherically-symmetric sources, in addition to two other methods: the one based on the series expansion of the Euler kernel, Eq. (\ref{eq8.12}), outlined in Appendix \ref{sectiongalacticappa}, and the direct computation of the gravitational field, Eq. (\ref{eq3.8}), outlined at the end of this section. We will not use this third method in Eq. (\ref{eq3.6}) for our galactic fits, but we checked that it yields results equivalent to the other methods, in some practical cases. 
	\par
}

It is not possible to simplify Eq. (\ref{eq3.6}) for a non-zero spherical mass distribution in the range $R_{1} <r^{ \prime } <R_{2}$,  and obtain expressions for the inner cavity ($r^{ \prime } <R_{1}$) and the outer region ($r^{ \prime } >R_{2}$) similar to those in  Eq. (\ref{eq3.5}). However, it is possible to consider the case of a constant mass distribution $\rho \left (r^{ \prime }\right ) =\rho _{0}$ inside the shell ($R_{1} <r^{ \prime } <R_{2}$) and zero outside, and then integrate Eq. (\ref{eq3.6}) between $R_{1}$ and $R_{2}$.

For example, if the fractional dimension $D$ is related to just one of the three $\alpha _{i}$ parameters, i.e. $D =\alpha _{i} +2$, in the inner cavity ($r^{ \prime } <R_{1}$) and in the outer region ($r^{ \prime } >R_{2}$) we obtain, respectively:
\begin{align}\Phi _{in}\left (r\right ) = -\frac{2\pi G\rho _{0}\left (R_{2}^{2} -R_{1}^{2}\right )}{\left (D -2\right )l_{0}} \label{eq3.7} \\
	\Phi _{out}\left (r\right ) = -\frac{4\pi G\rho _{0}\left (R_{2}^{D} -R_{1}^{D}\right )}{D\left (D -2\right )l_{0}r^{D -2}} , \nonumber \end{align}showing that inside the cavity the gravitational potential is constant and, therefore, the field is zero (no EFE), while the standard NFDG $1/r^{D -2}$ behavior is recovered outside the spherical shell. Similar results can be obtained when the overall dimension $D$ depends on more than one parameter $\alpha _{i}$. Although we have only considered in Eq. (\ref{eq3.7}) the case of a constant mass distribution $\rho _{0}$, which allows for a simple integration of Eq. (\ref{eq3.6}), these results are applicable to a thin spherical shell, whose mass distribution can be considered practically constant within the shell, thus proving that NFDG does not imply the EFE or a violation of the SEP.

A more general proof of the same statement was already provided in our paper I. In Appendix B of Ref. \cite{Varieschi:2020ioh}, it was shown how to derive a general expression of the NFDG gravitational field
$\mathbf{g}(r)$, for a spherically-symmetric mass distribution $\rho \left (r^{ \prime }\right )$, in a space of fractional dimension
$1 \leq D(r^{}) \leq 3$:

\begin{equation}\mathbf{g}_{NFDG}(r) = -\frac{4\pi G}{l_{0}^{2}r^{D\left (r\right )^{} -1}}{\displaystyle\int \nolimits_{0}^{r}}\rho \left (r^{ \prime }\right )r^{ \prime ^{D\left (r\right )^{} -1}}dr^{ \prime }\overset{}{\widehat{\mathbf{r}} ,} \label{eq3.8}
\end{equation}
which reduces to the Newtonian expression for fixed dimension
$D =3$: 

\begin{equation}\mathbf{g}_{Newt}(r) = -\frac{4\pi G}{l_{0}^{2}r^{2}}{\displaystyle\int \nolimits_{0}^{r}}\rho \left (r^{ \prime }\right )r^{ \prime ^{2}}dr^{ \prime }\overset{}{\widehat{\mathbf{r}} .} \label{eq3.9}
\end{equation}

These two fields were then identified respectively with the
``observed'' field
$\mathbf{g}_{obs}$ and the ``baryonic'' field
$\mathbf{g}_{bar}$, and used successfully to model several spherically-symmetric mass distributions, which were analyzed in paper I. Eq. (\ref{eq3.8}) was obtained by expanding the Newtonian derivation found in standard textbooks \cite{fowles2005analytical}, based on the computation of the field due to an infinitesimal spherical shell (see paper I\ for full details). The total field $\mathbf{g}_{NFDG}\left (r\right )$ at radial distance $r$, was obtained by integration over all ``inner'' shells, considering different geometries for the cases $2 <D <3$, $D =2$, $1 <D <2$, $D =1$, while it was shown that ``outer'' shell contributions are identically zero in all these cases. Therefore, in NFDG, Newton's shell theorem for a spherical distribution of matter still holds: \textit{a body that is inside a spherical shell of matter experiences no net gravitational force from that shell}.\protect\footnote{
	The only case in NFDG where Newton's shell theorem does not seem to hold is for $0 <D <1$ \cite{Varieschi:2020ioh}, where the outer ``shells'' yield non-zero contributions to the field. Thus, the EFE and the SEP violation might be possible for low values of the fractional dimension $D$. However, all galactic structures analyzed with our methods typically show $D >1$, and as a consequence they are not affected by the EFE.
}

From all the previous arguments, we conclude that in NFDG there should not be any violation of the SEP, in relation to possible detection of the EFE for galactic structures whose variable fractional dimension is in the range $1 \leq D \leq 3$. In the next section, we will present new results for four different galaxies, which do not seem to show any evidence of a possible external field effect, at least within the limits of the NFDG model.

\section{Galactic data fitting}
\label{sectiongalactic}

In the following subsections, we will apply NFDG to four notable examples of rotationally supported galaxies from the SPARC database: NGC 5033, NGC 6674, NGC 5055, and NGC 1090. The choice of these galaxies is due to the fact that they were used as the main examples for testing the Strong Equivalence Principle and detecting the EFE in the recent papers by Chae et al. \cite{Chae:2020omu,Chae:2021dzt} on the subject. The detailed luminosity data for the bulge, disk, and gas components of all these galaxies were obtained from the publicly available SPARC data \cite{Lelli:2016zqa}, supplemented by additional information from the database administrator \cite{Lelli:2021pri}.

\subsection{\label{sectiongalacticone}NGC 5033}
\label{sectiongalacticone}

We will start with the case of NGC 5033, a spiral galaxy located in the constellation Canes Venatici and approximately 38-60 million light-years away. This galaxy has a very bright central nucleus and a relatively faint disk extending outward in space. This is evidenced by the SPARC luminosity data, which show a strong bulge component up to about $1\ kpc$\textit{\textrm{}}\textrm{}, beyond which the stellar disk dominates up to the largest radial distances, together with a relatively less strong gas component.

SPARC data for this galaxy include the following:\ distance $D =\left (15.70 \pm 4.70\right )$ \textrm{Mpc}, disk scale length $R_{d} =5.16\ kpc =1.59 \times 10^{20}\mbox{m}$, asymptotically flat rotation velocity $V_{f} =\left (194.2 \pm 3.6\right )\mbox{km}/\mbox{s}$. From the SPARC luminosity data for the three main components, we obtained the corresponding volume and surface mass distributions, following the procedure discussed in Appendix \ref{sectiongalacticappa}. By integrating these mass distributions we computed the following galactic masses: $M_{bulge} =3.85 \times 10^{40}\mbox{kg}$, $M_{disk} =8.67 \times 10^{40}\mbox{kg}$, $M_{gas} =2.84 \times 10^{40}\mbox{kg}$, and $M_{total} =1.54 \times 10^{41}\mbox{kg}$.

\begin{figure}\centering 
	\setlength\fboxrule{0in}\setlength\fboxsep{0.1in}\fcolorbox[HTML]{000000}{FFFFFF}{\includegraphics[ width=6.99in, height=8.728805970149253in,]{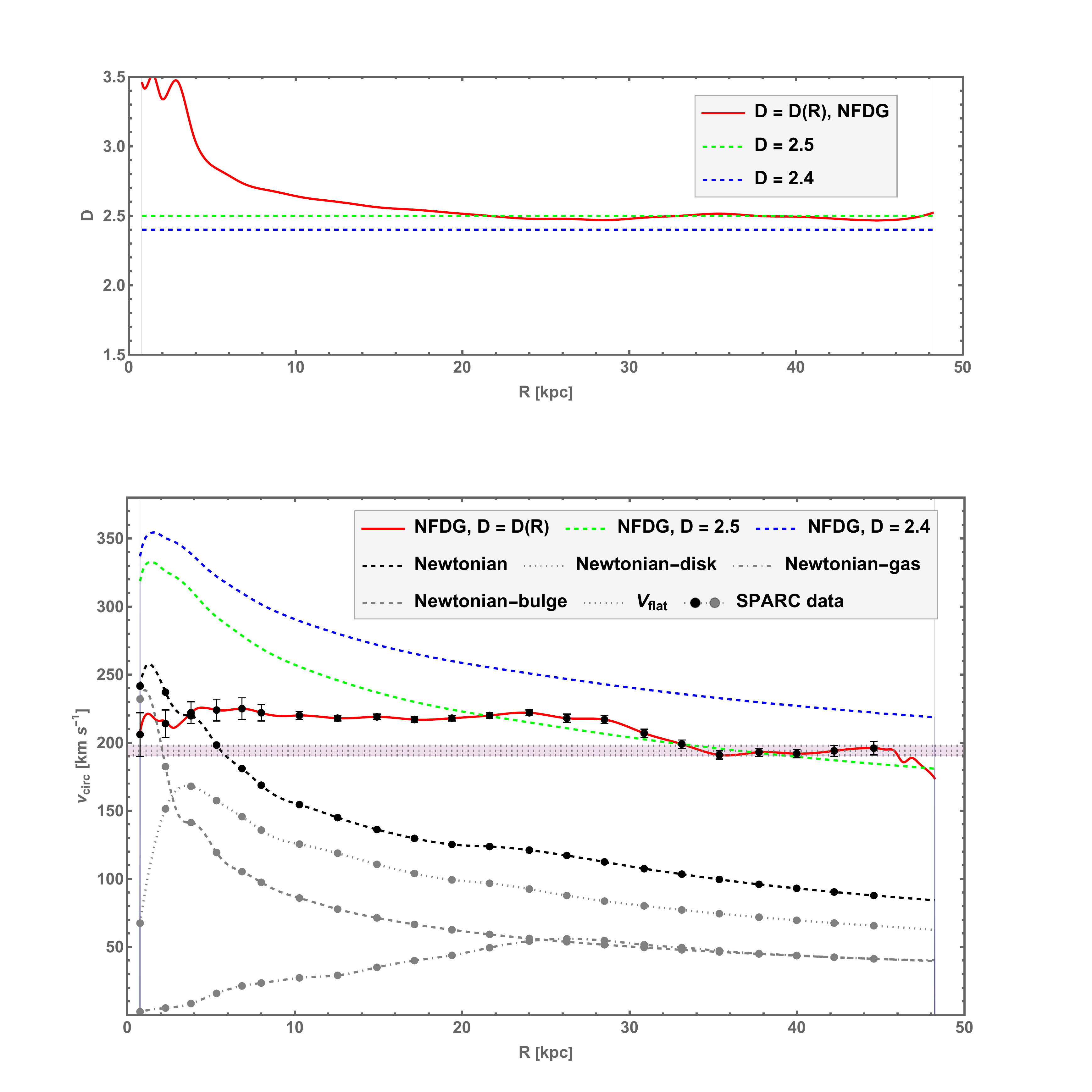}}
	\caption{NFDG results for NGC 5033.
		Top panel: NFDG variable dimension $D\left (R\right )$, based directly on SPARC data (red-solid curve), compared with fixed values $D =2.5$ and $D =2.4$ (green and blue-dashed lines). Bottom panel: NFDG rotation curves (circular velocity vs. radial distance) compared to the original SPARC data (black circles with error bars). The NFDG best fit for the variable dimension $D\left (R\right )$ is shown by the red-solid line, while NFDG fits for fixed values $D =2.5$ and $D =2.4$ are shown respectively by the green and blue dashed lines. Also shown: Newtonian rotation curves (different components - gray lines, total - black dashed line) with corresponding SPARC data (gray/black circles), and asymptotic flat velocity band (horizontal gray band).}
	\label{figure:NGC5033_1}
\end{figure}

The NFDG results for this galaxy are illustrated in Fig. \ref{figure:NGC5033_1}. As already done in our previous paper III, we will use the astrophysical radial distance $R$ measured in \textrm{kiloparsec}, while rotation circular velocities $v_{circ}$ will be measured in $\mbox{km}\ \mbox{s}^{ -1}$. The computations were performed with the methods detailed in Appendix \ref{sectiongalacticappa}, with the radial limits of the computation set at $R_{\min } =0.76\ \ kpc
$ and $R_{\max } =48.2\ \ kpc$. These limits are shown in all figures as vertical thin-gray lines.

The main NFDG result in this figure is the variable dimension $D\left (R\right )$ shown in the top panel by the red-solid curve. As in our previous papers, this was obtained by interpolating the experimental SPARC data for the circular velocities (black circles with error bars in the bottom panel) and obtaining from them the equivalent observed gravitational field $g_{obs}\left (R\right )$, based solely on SPARC experimental data. We then assumed that in NFDG the observed field is due to $g_{NFDG}\left (R\right )$ as in Eq. (\ref{eq2.8}), obtained by computing the NFDG gravitational potentials as described in Appendix \ref{sectiongalacticappa}.

These potentials and related gravitational field can be considered, at each field point, also as a function of the variable dimension $D\left (R\right )$, i.e., $g_{NFDG}\left (R ,D\left (R\right )\right )$. The computing range $\left (R_{\min } ,R_{\max }\right )$ was then subdivided into one hundred equal sub-intervals with related radial distances $R_{i} =R_{\min } +i\genfrac{(}{)}{}{}{R_{\max } -R_{\min }}{100} ,\ \ i =0 ,... ,100$, and for each of these points $R_{i}$ we solved numerically the following equation:

\begin{equation}g_{NFDG}\left (R_{i} ,D\left (R_{i}\right )\right ) =g_{obs}\left (R_{i}\right ) , \label{eq4.1}
\end{equation}
in order to determine the corresponding value of the variable dimension $D_{i} \equiv D\left (R_{i}\right ) ,i =1 ,... ,100$.

By interpolating this set of $\left \{\left (R_{i} ,D_{i}\right )\right \}$ points, we then obtained the main NFDG red-solid variable dimension $D\left (R\right )$ curve in the top panel of Fig. \ref{figure:NGC5033_1}. To double-check our calculation, for each radial point $R_{i}$ we recomputed the NFDG circular velocities using the $D\left (R\right )$ curve and Eq. (\ref{eq2.9}), and obtained the perfect NFDG fit to the SPARC experimental data shown by the red-solid curve in the bottom panel of Fig. \ref{figure:NGC5033_1}. Again, this perfect agreement is expected, since at each point we selected the appropriate value of the variable dimension $D\left (R_{i}\right )$ which is able to match the experimental value $g_{obs}\left (R_{i}\right )$ with the predicted NFDG value $g_{NFDG}\left (R_{i} ,D\left (R_{i}\right )\right )$, following Eq. (\ref{eq4.1}).

In our previous papers, we used also to compare NFDG computations with the general RAR relation and derive an additional variable dimension function based on this more general relation, together with a corresponding circular velocity fit for each galaxy. In the current paper, we don't feel that this is important anymore, since in NFDG each galaxy is characterized by its own fractional dimension function and, therefore, we have omitted this additional analysis in the current work.

However, in Fig. \ref{figure:NGC5033_1} we also show the NFDG circular velocity fits for two particular fixed values of the variable dimension: $D =2.5$ (green-dashed curves) and $D =2.4$ (blue-dashed curves). These values were selected by looking at the variable dimension $D =D\left (R\right )$ in the top panel (red curve) and by choosing two fixed values for $D$ which were close enough to the red curve in the asymptotic regime at larger radial distances. The corresponding NFDG circular velocity curves in the bottom panel show that these fixed value curves can only partially fit the experimental data (for example, the $D =2.5$ green-dashed curve is a reasonably good fit at larger distances for this galaxy, but not at lower distances), thus confirming that a variable fractional dimension function is indeed required in NFDG to obtain good fits of the rotation curves. 

Comparison of our main NFDG results (red-solid curves) with the two fixed-$D$ curves (green-dashed and blue-dashed curves) can also be used for a rough estimate of the uncertainty in the variable dimension $D(R)$, which is of the order $\Delta D\lesssim 0.1$ over most of the radial range, as it will be shown in more detail in Appendix \ref{sectiongalacticdeep5055}. In the following subsections, we will perform a similar analysis for all the other galaxies considered in this work.

In the bottom panel of Fig. \ref{figure:NGC5033_1}, we also show for completeness the Newtonian rotation curves (different components - gray lines; total - black dashed line) with corresponding SPARC data (gray/black circles), and the asymptotic flat velocity band (horizontal gray band), based on the SPARC data for this galaxy reported above. These Newtonian curves are obtained directly by interpolating the original SPARC data, but we also checked the consistency of our NFDG\ potentials and fields by computing the ``Newtonian'' case corresponding to a fixed $D =3$ dimension in all NFDG formulas. These results were in agreement with the Newtonian curves based on SPARC data, but they will not be reported in this work.

As a general remark about our NFDG results, we reiterate that the main conclusion is illustrated by the red-solid variable dimension curve in the top panel of Fig. \ref{figure:NGC5033_1}: if this galactic structure really behaves as a fractal object whose Hausdorff fractional dimension follows our NFDG $D\left (R\right )$ red-solid curve, then the observed rotation curve will just be a consequence of the NFDG field equations, without requiring any form of dark matter. The NFDG $D\left (R\right )$ curve for NGC 5033 seems to be close to standard $D \approx 3$ values at lower radial distances, where the spherical bulge is more dominant, while progressively decreasing at larger distances and approaching an almost constant $D \approx 2.5$ value at the largest radii, where the asymptotic flat velocity regime takes place. Similar behavior will be seen for the other galaxies presented in this paper, such as NGC 6674, NGC 5055, and NGC 1090.

The only unexpected feature of these results for NGC 5033 is the increase of the variable dimension, $D \approx 3.0 -3.5$, at the lowest radial distances. While we fixed previous problems with the convergence of our series at low radial distances (see Appendix \ref{sectiongalacticappa}), the NFDG calculations might still yield unphysical results at these lowest radii, also in view of lower precision in the mass distribution data for the central bulge component and related uncertainties. This particular low-$R$, $D \gtrsim 3$ behavior is also present in some of the other galaxies studied in this paper, and at the moment we are unable to explain this increase of the spatial dimension beyond the standard $D \approx 3$.

With regard to the EFE, NGC 5033 is considered by Chae et al. \cite{Chae:2020omu,Chae:2021dzt} as one of the ``golden galaxies'' for the EFE\ to be detected, since it is located in an exceptionally dense environment, and thus very susceptible to the effect of a strong local external gravitational field. As remarked instead in Sect. \ref{sectiongalacticefe} above, NFDG should not be affected by external gravitational fields, as in the case of standard Newtonian gravity. Therefore, we can only compare NFDG results for NGC 5033 (considered a MOND-EFE galaxy) with similar NFDG results for MOND-non-EFE galaxies, such as NGC 6674, which will be studied in the next subsection, in order to see possible differences between the EFE and non-EFE cases.

\subsection{NGC 6674}
\label{sectiongalactictwo}

As our second galaxy we consider here NGC 6674, a barred spiral galaxy in the constellation Hercules. Similar to the previous case analyzed in Sect. \ref{sectiongalacticone}, this galaxy also possesses a strong bulge component up to about $1 -2\ kpc$, beyond which the stellar disk component dominates up to the largest radial distances, together with a relatively less strong gas component.

\begin{figure}\centering 
	\setlength\fboxrule{0in}\setlength\fboxsep{0.1in}\fcolorbox[HTML]{000000}{FFFFFF}{\includegraphics[ width=6.99in, height=8.728805970149253in,]{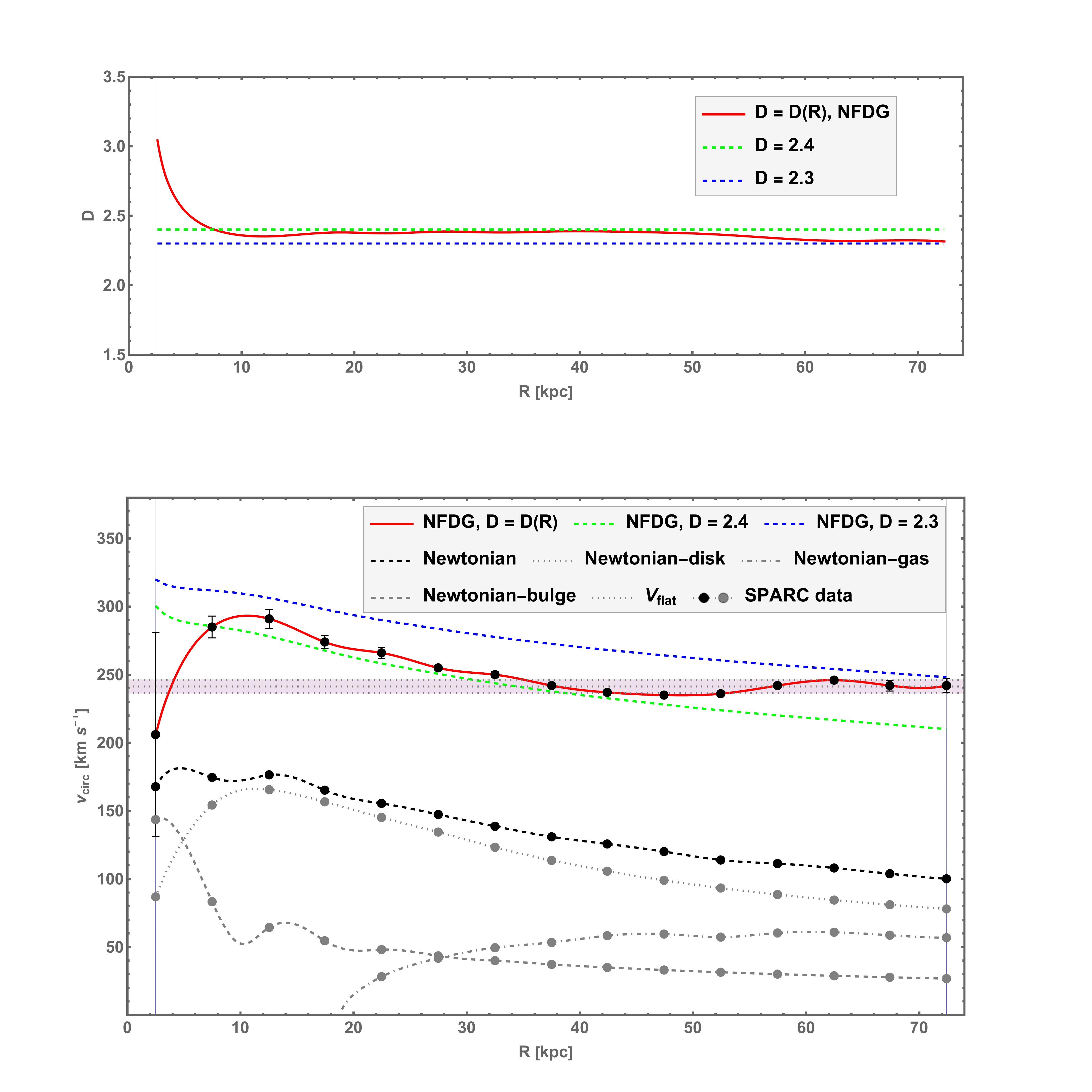}
	}
	\caption{NFDG results for NGC 6674.
		Top panel: NFDG variable dimension $D\left (R\right )$, based directly on SPARC data (red-solid curve), compared with fixed values $D =2.4$ and $D =2.3$ (green and blue-dashed lines). Bottom panel: NFDG rotation curves (circular velocity vs. radial distance) compared to the original SPARC data (black circles with error bars). The NFDG best fit for the variable dimension $D\left (R\right )$ is shown by the red-solid line, while NFDG fits for fixed values $D =2.4$ and $D =2.3$ are shown respectively by the green and blue dashed lines. Also shown: Newtonian rotation curves (different components - gray lines, total - black dashed line) with corresponding SPARC data (gray/black circles), and asymptotic flat velocity band (horizontal gray band).}
	\label{figure:NGC6674_1}
\end{figure}

SPARC data for this galaxy include the following:\ distance $D =\left (51.20 \pm 10.20\right )$ \textrm{Mpc}, disk scale length $R_{d} =6.04\ kpc =1.86 \times 10^{20}\mbox{m}$, asymptotically flat rotation velocity $V_{f} =\left (241.3 \pm 4.9\right )\mbox{km}/\mbox{s}$. By integrating the SPARC mass distributions we computed the following galactic masses: $M_{bulge} =4.35 \times 10^{40}\mbox{kg}$, $M_{disk} =2.04 \times 10^{41}\mbox{kg}$, $M_{gas} =7.74 \times 10^{40}\mbox{kg}$, and $M_{total} =3.25 \times 10^{41}\mbox{kg}$. The NFDG results for this galaxy are illustrated in Fig. \ref{figure:NGC6674_1}, with the radial limits set at $R_{\min } =2.48\ \ kpc
$ and $R_{\max } =72.4\ \ kpc$ (vertical thin-gray lines in the figure).

The variable dimension $D\left (R\right )$ is shown in the top panel by the red-solid curve, and was obtained with the same procedures following Eq. (\ref{eq4.1}) above.  The NFDG circular velocities were then computed using this $D\left (R\right )$ curve and Eq. (\ref{eq2.9}), producing a perfect NFDG fit to the SPARC experimental data shown by the red-solid curve in the bottom panel of Fig. \ref{figure:NGC6674_1}.

In Fig. \ref{figure:NGC6674_1}, we also show the NFDG circular velocity fits for two fixed values of the variable dimension: $D =2.4$ (green-dashed curves) and $D =2.3$ (blue-dashed curves). These fixed values were the closest to the red curve in the asymptotic regime at larger radial distances in the top panel. The corresponding NFDG circular velocity curves, in the bottom panel, show that these fixed value curves can partially fit the experimental data (for example, the $D =2.3$ blue-dashed curve is a close fit at the largest distances for this galaxy, but not at lower distances, where the $D =2.4$ green-dashed curve is actually a better fit).

In the bottom panel of Fig. \ref{figure:NGC6674_1}, we also show for completeness the additional information which will also be included in all the other figures of this paper (Newtonian rotation curves with corresponding SPARC data, asymptotic flat velocity band, etc.). The overall analysis for this galaxy is very similar to the previous one for NGC 5033: these two galaxies have similar mass distributions in terms of the three components (bulge, disk, gas); the NFDG $D\left (R\right )$ curve for NGC 6674 is also close to standard $D \approx 3$ values at lower radial distances where the spherical bulge is more dominant, while progressively decreases at larger distances and approaches an almost constant $D \approx 2.3$ value at the largest radii, where the asymptotic flat velocity regime takes place. At the lowest radial distances, the variable dimension seems to become $D \gtrsim 3$, but this could be just an (unphysical) effect of our simulations at low $R$.

In the analysis by Chae et al. \cite{Chae:2020omu,Chae:2021dzt}, NGC 6674 is considered an exceptionally isolated galaxy, and a perfect example of a ``control galaxy'' for null detection of the EFE. Since NFDG does not expect any EFE, we can only compare our results for NGC 6674 with the case of NGC 5033, which is similar in terms of mass distributions and other galactic parameters (see also the case of NGC 7814, later analyzed in Appendix \ref{sectiongalacticfive}). Comparing Fig. \ref{figure:NGC6674_1} with Fig. \ref{figure:NGC5033_1} (and also with Fig. \ref{figure:NGC7814_1} in Appendix \ref{sectiongalacticfive}), we don't notice any peculiar differences between all these NFDG results. 

In particular, we do not observe any apparent difference in the slope of the variable dimension functions at the largest radial distances (around the last few experimental data points) for these comparable galaxies: these $D(R)$ functions appear to have similar slopes, slightly decreasing, in the region within the last four experimental data points. As it will be discussed in Appendix \ref{sectiongalacticdeep5055}, a difference in the slope of the dimension function in the outskirt of the plot might be an indication of the EFE: in particular, a slightly increasing slope might indicate the EFE, while a slightly decreasing slope is more likely to be associated with a non-EFE case.

\subsection{NGC 5055}
\label{sectiongalacticthree}

The third galaxy in our analysis is NGC 5055, a spiral galaxy in the constellation of Canes Venatici, also known as M63 or the \textit{Sunflower Galaxy}. This is also considered \cite{Chae:2020omu,Chae:2021dzt} a ``golden galaxy'' for the EFE detection, due to the dense galactic environment.

\begin{figure}\centering 
	\setlength\fboxrule{0in}\setlength\fboxsep{0.1in}\fcolorbox[HTML]{000000}{FFFFFF}{\includegraphics[ width=6.99in, height=8.728805970149253in,]{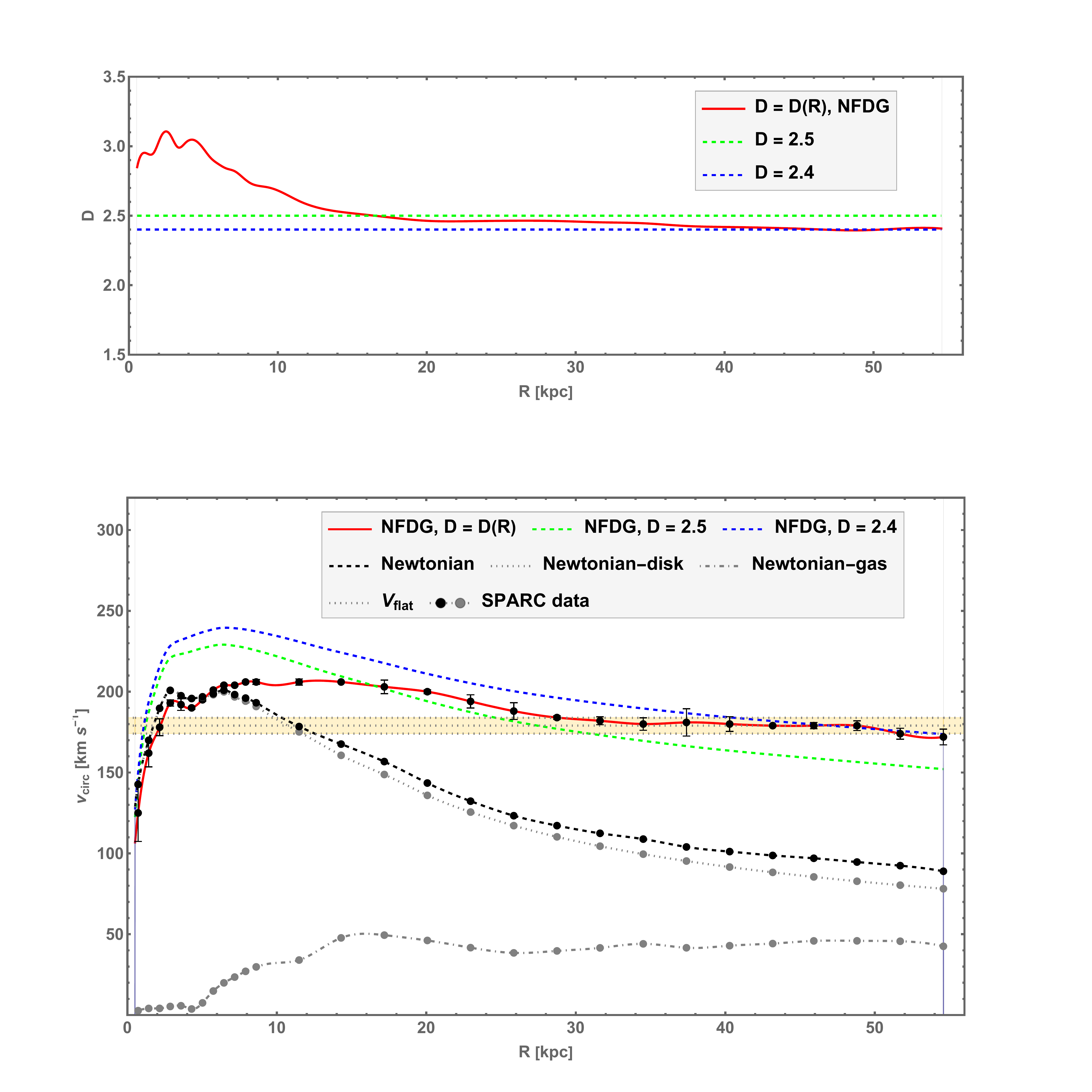}
	}
	\caption{NFDG results for NGC 5055.
		Top panel: NFDG variable dimension $D\left (R\right )$, based directly on SPARC data (red-solid curve), compared with fixed values $D =2.5$ and $D =2.4$ (green and blue-dashed lines). Bottom panel: NFDG rotation curves (circular velocity vs. radial distance) compared to the original SPARC data (black circles with error bars). The NFDG best fit for the variable dimension $D\left (R\right )$ is shown by the red-solid line, while NFDG fits for fixed values $D =2.5$ and $D =2.4$ are shown respectively by the green and blue dashed lines. Also shown: Newtonian rotation curves (different components - gray lines, total - black dashed line) with corresponding SPARC data (gray/black circles), and asymptotic flat velocity band (horizontal gray band).}
	\label{figure:NGC5055_1}
\end{figure}

SPARC data for this galaxy only show disk and gas components, with the stellar disk as the dominating one. Also from SPARC data for this galaxy we have the following:\ distance $D =\left (9.90 \pm 0.30\right )$ \textrm{Mpc}, disk scale length $R_{d} =3.20\ kpc =9.88 \times 10^{19}\mbox{m}$, asymptotically flat rotation velocity $V_{f} =\left (179.0 \pm 4.9\right )\mbox{km}/\mbox{s}$. By integrating the SPARC mass distributions we computed the following galactic masses: $M_{disk} =1.48 \times 10^{41}\mbox{kg}$, $M_{gas} =3.12 \times 10^{40}\mbox{kg}$, and $M_{total} =1.79 \times 10^{41}\mbox{kg}$. The NFDG results for this galaxy are illustrated in Fig. \ref{figure:NGC5055_1}, with the radial limits set at $R_{\min } =0.512\ \ kpc
$ and $R_{\max } =54.6\ \ kpc$ (vertical thin-gray lines in the figure).

The variable dimension $D\left (R\right )$ is shown in the top panel as a red-solid curve, while NFDG circular velocities were computed using this $D\left (R\right )$ curve and Eq. (\ref{eq2.9}), producing again a perfect NFDG fit to the SPARC experimental data shown by the red-solid curve in the bottom panel of Fig. \ref{figure:NGC5055_1}. As for the previous galaxies, in Fig. \ref{figure:NGC5055_1} we also show the NFDG circular velocity fits for two fixed values of the variable dimension: $D =2.5$ (green-dashed curves) and $D =2.4$ (blue-dashed curves). The corresponding NFDG circular velocity curves in the bottom panel, show that these fixed value curves can partially fit the experimental data (for example, the $D =2.4$ blue-dashed curve is a very good fit at the largest distances for this galaxy, but not at lower distances).

The overall analysis for this galaxy is very similar to the previous two, although NGC 5055 does not have a central bulge component: the NFDG $D\left (R\right )$ curve for NGC 5055 is close to standard $D \approx 3$ values at lower radial distances, while progressively decreasing at larger distances and approaching an almost constant $D \approx 2.4$ value at the  largest radii, where the asymptotic flat velocity regime takes place. At the lowest radial distances, the variable dimension seems to exceed $D =3$, as already noted in previous cases. No particular NFDG feature seems to imply any EFE detection for this galaxy, which will be compared with a similar non-EFE galaxy in the next section and also analyzed in more detail in Appendix \ref{sectiongalacticdeep5055}.

\subsection{NGC 1090}
\label{sectiongalacticfour}

\begin{figure}\centering 
	\setlength\fboxrule{0in}\setlength\fboxsep{0.1in}\fcolorbox[HTML]{000000}{FFFFFF}{\includegraphics[ width=6.99in, height=8.728805970149253in,]{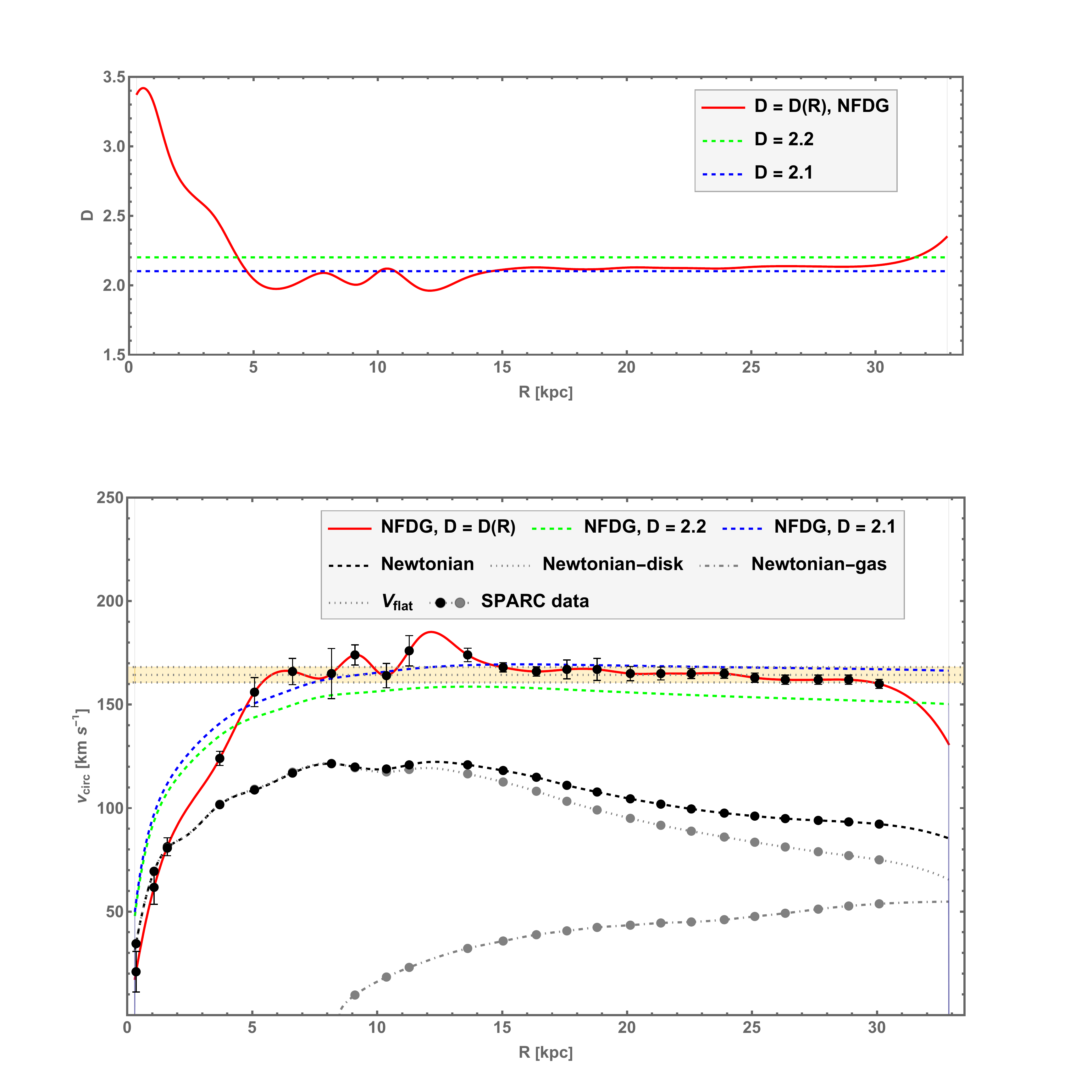}
	}
	\caption{NFDG results for NGC 1090.
		Top panel: NFDG variable dimension $D\left (R\right )$, based directly on SPARC data (red-solid curve), compared with fixed values $D =2.2$ and $D =2.1$ (green and blue-dashed lines). Bottom panel: NFDG rotation curves (circular velocity vs. radial distance) compared to the original SPARC data (black circles with error bars). The NFDG best fit for the variable dimension $D\left (R\right )$ is shown by the red-solid line, while NFDG fits for fixed values $D =2.2$ and $D =2.1$ are shown respectively by the green and blue dashed lines. Also shown: Newtonian rotation curves (different components - gray lines, total - black dashed line) with corresponding SPARC data (gray/black circles), and asymptotic flat velocity band (horizontal gray band).}
	\label{figure:NGC1090_1}
\end{figure}

The fourth galaxy in our analysis is NGC 1090, a barred spiral galaxy in the constellation Cetus. This is considered an exceptionally isolated non-EFE ``control galaxy'' in Refs. \cite{Chae:2020omu,Chae:2021dzt}. SPARC data for this galaxy only show disk and gas components, with the stellar disk as the dominating one, and also the following:\ distance $D =\left (37.00 \pm 9.25\right )$ \textrm{Mpc}, disk scale length $R_{d} =3.53\ kpc =1.09 \times 10^{20}\mbox{m}$, asymptotically flat rotation velocity $V_{f} =\left (164.4 \pm 3.7\right )\mbox{km}/\mbox{s}$. The computed galactic masses are: $M_{disk} =7.26 \times 10^{40}\mbox{kg}$, $M_{gas} =2.29 \times 10^{40}\mbox{kg}$, and $M_{total} =9.55 \times 10^{40}\mbox{kg}$. The NFDG results for this galaxy are illustrated in Fig. \ref{figure:NGC1090_1}, with the radials limits set at $R_{\min } =0.299\ \ kpc
$ and $R_{\max } =32.9\ \ kpc$.

In addition to the variable dimension $D\left (R\right )$ (top panel, red-solid curve) and the NFDG fit to the SPARC experimental data (red-solid curve, bottom panel), we also show the NFDG circular velocity fits for two fixed values of the variable dimension: $D =2.2$ (green-dashed curves) and $D =2.1$ (blue-dashed curves). In this case, the $D =2.1$ blue-dashed curve is a close fit at most distances for this galaxy, except at the lowest ones.

The overall analysis for this galaxy is very similar to NGC 5055 (or NGC 6503 in Appendix \ref{sectiongalacticfive}): the NFDG $D\left (R\right )$ curve for NGC 1090 is close to $D \gtrsim 3$ values at lower radial distances, while progressively decreasing at larger distances toward almost constant $D \approx 2.1 -2.2$ values at the  largest radii, where the asymptotic flat velocity regime takes place. Comparing this non-EFE galaxy with the previous NGC 5055 EFE galaxy, we do not notice any difference in the NFDG analysis, as already remarked for the previous cases. Again, as discussed in Appendix \ref{sectiongalacticdeep5055}, the similar slopes of the $D(R)$ curves for both NGC 5055 and NGC 1090 in the outskirt of the plot (region within the last few experimental points) should indicate a non-EFE situation for both galaxies.

\section{Conclusions}
\label{sectiongalacticconcl}

In this work, we continued investigating NFDG, a fractional-dimension gravity model which might yield a possible explanation of the galactic rotation curves, without any dark matter hypothesis. After reviewing the fundamental ideas of NFDG introduced in our papers I-IV, we analyzed the external field effect as a possible test to distinguish between standard Newtonian and Einstein gravity (non-EFE theories) and alternative models such as MOND, Conformal Gravity, and Modified Gravity (EFE theories). However, NFDG does not seem to imply any violation of the strong equivalence principle, by means of the external field effect and its experimental detection in galactic dynamics.

We extended our catalog of NFDG galactic rotation curves by studying  four more rotationally supported galaxies (NGC 5033, NGC 6674, NGC 5055, NGC 1090), and improved on our previous analysis of three other galaxies (NGC 7814, NGC 6503, NGC 3741 - Appendix \ref{sectiongalacticfive}). In particular, the first four galaxies were used by Chae et al. \cite{Chae:2020omu,Chae:2021dzt} in their recent MOND-EFE studies, so it was an obvious choice to consider them also within the framework of NFDG.

The general outcome of our NFDG analysis of these seven galaxies is that each one is characterized by a specific variable dimension function $D\left (R\right )$, with some common features: at low radial distances we typically have $D \approx 3$ (except for the lowest radii, with $D \gtrsim 3$ or $D \lesssim 2$, probably an unphysical result due to the limitations of our numerical computations). The variable dimension then progressively decreases toward an almost constant value ($D \approx 2.3 -2.5$, for most cases) at the largest radial distances, where the rotation curves achieve their flat velocity regime. Therefore, if these galaxies effectively behave as fractal objects with Hausdorff fractional dimension following their respective $D\left (R\right )$ functions, then the observed rotation curves will be a simple consequence of the NFDG field equations, and no dark matter contribution of any kind will be required.

With regard to the EFE, we did not observe any particular difference in our NFDG\ experimental fits, between galaxies which are supposed to exhibit the EFE (NGC 5033, NGC 5055) and those which are considered to be non-EFE (NGC 6674, NGC 1090) within the MOND paradigm. Since NFDG does not imply any SEP violation through the external field effect, this result was expected. However, more galaxies will need to be analyzed with our methods before we can reach a final conclusion on the matter.

\section*{Acknowledgements}

This work was supported by the Seaver College of Science and Engineering, Loyola Marymount University - Los Angeles. The author also wishes to acknowledge Dr. Federico Lelli for sharing SPARC galactic data files and other useful information.




\appendix

\section{NFDG computations}
\label{sectiongalacticappa}

In this section, we will give full details of the computation of the NFDG potentials for the practical case of galactic structures. These details were already discussed in our papers I-III, but we prefer to summarize them again here, also in view of some minor changes in the computations.
Combining together Eqs. (\ref{eq2.5}) and (\ref{eq2.7}), the NFDG gravitational potential $\Phi _{NFDG}\left (\mathbf{x}\right )$
becomes:
\begin{gather}\Phi _{NFDG}(\mathbf{x}) = -\frac{2\pi ^{1 -D/2}\Gamma (D/2)G}{\left (D -2\right )l_{0}}{\displaystyle\int \nolimits_{\mathcal{V}}}d^{3}\mathbf{x}^{ \prime }v\left (\mathbf{x}^{ \prime }\right )\frac{\rho (\mathbf{x}^{ \prime })}{\left \vert \mathbf{x} -\mathbf{x}^{ \prime }\right .\vert ^{D -2}};\ D \neq 2 \label{eq8.1} \\
	\Phi _{NFDG}\left (\mathbf{x}\right ) =\frac{2G}{l_{0}}{\displaystyle\int \nolimits_{\mathcal{V}}}d^{2}\mathbf{x}^{ \prime }v\left (\mathbf{x}^{ \prime }\right )\rho \left (\mathbf{x}^{ \prime }\right )\ln \left \vert \mathbf{x} -\mathbf{x}^{ \prime }\right .\vert ;\ D =2 , \nonumber \end{gather}
where, in general, $D =D\left (\mathbf{x}\right )$ for the case $D \neq 2$, but with a slowly varying fractional dimension so that we can ignore space derivatives of this function when computing the NFDG gravitational field: $\mathbf{g}_{NFDG}\left (\mathbf{x}\right ) = - \nabla \Phi _{NFDG}\left (\mathbf{x}\right )/l_{0}$.

As previously mentioned, the mass distributions of the galaxies being studied typically have one spherically-symmetric component (galactic bulge) and two components of cylindrical symmetry (stellar disk and gas disk components). We will use (dimensionless) spherical coordinates ($r$, $\theta $, $\phi $) for the first component and (dimensionless) cylindrical coordinates ($R$, $\phi $, $z$) for the other two components.

The main element of the general NFDG potential in the first line of Eq. (\ref{eq8.1}), is the so-called Euler kernel, $1/\left \vert \mathbf{x} -\mathbf{x}^{ \prime }\right \vert ^{D -2}$, which admits the following multipole expansion for $D >1$, $D \neq 2$ in (rescaled) spherical coordinates \cite{2012JPhA...45n5206C}:

\begin{equation}\frac{1}{\left \vert \mathbf{x} -\mathbf{x}^{ \prime }\right \vert ^{D -2}} =\sum \limits _{l =0}^{\infty }\frac{r_{ <}^{l}}{r_{ >}^{l +D -2}}C_{l}^{\left (\frac{D}{2} -1\right )}\left (\cos \gamma \right ) , \label{eq8.2}
\end{equation}
where $r_{ <}$ ($r_{ >}$) is the smaller (larger) of $r$ and $r^{ \prime }$, $\gamma $ is the angle between the unit vectors $\widehat{\mathbf{r}}$ and $\widehat{\mathbf{r}}^{ \prime }$, and $C_{l}^{\left (\lambda \right )}\left (x\right )$ denotes Gegenbauer polynomials (see paper I or Ref. \cite{NIST:DLMF} for general properties of these special functions).

Expansion (\ref{eq8.2}) can be used directly to model spherically-symmetric galactic components (bulge component) by aligning the field point vector $\mathbf{r}$ in the direction of the $z^{ \prime }$ axis and then using standard spherical coordinates ($r^{ \prime }$, $\theta ^{ \prime }$, $\phi ^{ \prime }$) for the source point vector $\mathbf{r}^{ \prime }$. In this way, we can use expansion (\ref{eq8.2})  with $\theta ^{ \prime }$ replacing the angle $\gamma $. The same expansion can also be applied to the case of cylindrical coordinates ($R_{}$, $\phi $, $z_{}$), following the analysis in paper II. In the case of thin galactic disks, in the $z =z^{ \prime } =0$ plane and in the $\phi  =0$ direction, the angle $\gamma $ is replaced by $\phi ^{ \prime }$ and the radial (dimensionless) spherical coordinate $r$ with the radial (dimensionless) cylindrical $R$:
\begin{equation}\frac{1}{\left \vert \mathbf{x} -\mathbf{}\mathbf{x}^{ \prime }\right \vert ^{D -2}} =\sum \limits _{l =0}^{\infty }\frac{R_{ <}^{l}}{R_{ >}^{l +D -2}}C_{l}^{\left (\frac{D}{2} -1\right )}\left (\cos \phi ^{ \prime }\right ) , \label{eq8.3}
\end{equation}
while, for thick galactic disks with $z^{ \prime } \neq 0$ (but still in the $z_{} =0$ plane and $\phi  =0$ direction), the following coordinate transformation can be used to modify the expansion (\ref{eq8.2}):
\begin{gather}r_{} =R_{} \label{eq8.4} \\
	r^{ \prime } =\sqrt{R^{ \prime 2} +z^{ \prime 2}} \nonumber  \\
	\cos \gamma  =\frac{R^{ \prime }\cos \phi ^{ \prime }}{\sqrt{R^{ \prime 2} +z^{ \prime 2}}} . \nonumber \end{gather}

The other key element of the general NFDG potential in the first line of Eq. (\ref{eq8.1}) is the weight $v\left (\mathbf{x}^{ \prime }\right )$, which follows from the original theory for spaces of non-integer dimension by Stillinger \cite{doi:10.1063/1.523395}, Svozil \cite{1987JPhA...20.3861S}, Palmer and Stavrinou \cite{Palmer_2004}, or from more recent theories of multi-scale spacetimes and fractional gravity \cite{Calcagni:2011sz,Calcagni:2009kc,Calcagni:2010bj,Calcagni:2013yqa,Calcagni:2020ads,Calcagni:2021ipd,Calcagni:2021aap,Calcagni:2021mmj} (see also paper IV for a more detailed discussion).

Following our previous papers, we will consider the NFDG weight $v\left (\mathbf{x}^{ \prime }\right )$ in rectangular coordinates $x^{ \prime i} =\left (x^{ \prime } ,y^{ \prime } ,z^{ \prime }\right )$ as:
\begin{equation}v\left (\mathbf{x}^{ \prime }\right ) =\prod \limits _{i =1}^{3}\frac{\pi ^{\alpha _{i}/2}}{\Gamma \left (\alpha _{i}/2\right )}\left \vert x^{ \prime i}\right \vert ^{\alpha _{i} -1} , \label{eq8.5}
\end{equation}where $0 <\alpha _{i} \leq 1$ and $D =\alpha _{1} +\alpha _{2} +\alpha _{3}$ is the overall Hausdorff dimension of the space. The weight in the previous equation is factorizable in the coordinates, and reduces to unity for $\alpha _{1} =\alpha _{2} =\alpha _{3} =1$, i.e., $D =3$. 

When $D <3$, the fractional dimension can be assigned to just one of the three coordinates (for example, $\alpha _{1} =\alpha _{2} =1$, $\alpha _{3} =D -2$), or distributed over two or all three coordinates (for example, distributed evenly over the three coordinates $\alpha _{1} =\alpha _{2} =\alpha _{3} =D/3$).

In spherical coordinates ($r^{ \prime }$, $\theta ^{ \prime }$, $\phi ^{ \prime }$), we can use standard coordinate transformations ($x^{ \prime } =r^{ \prime }\sin \theta ^{ \prime }\cos \phi ^{ \prime }$, $y^{ \prime } =r^{ \prime }\sin \theta ^{ \prime }\sin \phi ^{ \prime }$, $z^{ \prime } =r^{ \prime }\cos \theta ^{ \prime }$) to transform the NFDG weight into the following:

\begin{equation}v\left (\mathbf{x}^{ \prime }\right ) =\left (\prod \limits _{i =1}^{3}\frac{\pi ^{\alpha _{i}/2}}{\Gamma \left (\alpha _{i}/2\right )}\right )r^{ \prime \alpha _{1} +\alpha _{2} +\alpha _{3} -3}\left \vert \sin \theta ^{ \prime }\right \vert ^{\alpha _{1} +\alpha _{2} -2}\left \vert \cos \theta ^{ \prime }\right \vert ^{\alpha _{3} -1}\left \vert \sin \phi ^{ \prime }\right \vert ^{\alpha _{2} -1}\left \vert \cos \phi ^{ \prime }\right \vert ^{\alpha _{1} -1} , \label{eq8.6}
\end{equation}which will be combined with the standard volume element $d^{3}\mathbf{x}^{ \prime } =r^{ \prime 2}\sin \theta ^{ \prime }dr^{ \prime }d\theta ^{ \prime }d\phi ^{ \prime }$ inside the integrals in Eq. (\ref{eq8.1}). 

In our previous papers I-IV, we have already noted that there is a certain ambiguity in how to assign the fractional dimension to one or more of the three parameters $\alpha _{i}$. Since these parameters are originally related to the three rectangular coordinates, it seems more reasonable to spread the non-integer dimension equally over these three parameters, i.e., $\alpha _{1} =\alpha _{2} =\alpha _{3} =D/3$, rather than just one (or two) of them, so we have adopted this particular choice for the case of spherical symmetry.\protect\footnote{
	Other choices for the three parameters $\alpha _{i}$ were tested for some galactic structures, yielding results which were all very similar to each other. Therefore, the final results of our analysis do not depend much on the particular way the overall dimension $D$ is divided over the three $\alpha _{i}$ parameters.
}

Similarly, in cylindrical coordinates ($R^{ \prime }$, $\phi ^{ \prime }$, $z^{ \prime }$), we can use the coordinate transformations ($x^{ \prime } =R^{ \prime }\cos \phi ^{ \prime }$, $y^{ \prime } =R^{ \prime }\sin \phi ^{ \prime }$) to transform the NFDG weight into the following:

\begin{equation}v\left (\mathbf{x}^{ \prime }\right ) =\left (\prod \limits _{i =1}^{3}\frac{\pi ^{\alpha _{i}/2}}{\Gamma \left (\alpha _{i}/2\right )}\right )R^{ \prime \alpha _{1} +\alpha _{2} -2}\left \vert \sin \phi ^{ \prime }\right \vert ^{\alpha _{2} -1}\left \vert \cos \phi ^{ \prime }\right \vert ^{\alpha _{1} -1}\left \vert z^{ \prime }\right \vert ^{\alpha _{3} -1} , \label{eq8.7}
\end{equation}which will be combined with the standard volume element $d^{3}\mathbf{x}^{ \prime } =R^{ \prime }dR^{ \prime }d\phi ^{ \prime }dz^{ \prime }$ inside the integrals in Eq. (\ref{eq8.1}). Following our previous papers, for thin/thick galactic disks (for both gas and stellar disk components), we will simply set $\alpha _{3} =1$, i.e., no fractional dimension in the $z^{ \prime }$ direction, while we will divide evenly the fractional dimension over the other two parameters: $\alpha _{1} =\alpha _{2} =\frac{D -1}{2}$.

The last element of the general NFDG potential in Eq. (\ref{eq8.1}) is the (rescaled) mass density $\rho \left (\mathbf{x}^{ \prime }\right )$ which is obtained, for each galaxy, from the published SPARC data \cite{Lelli:2016zqa}. For thin/thick disk structures, the rescaled mass distributions are taken, respectively, as:
\begin{gather}\rho \left (R^{ \prime } ,z^{ \prime }\right ) =\Sigma \left (R^{ \prime }\right )\delta \left (z^{ \prime }\right ) \label{eq8.8} \\
	\rho \left (R^{ \prime } ,z^{ \prime }\right ) =\Sigma \left (R^{ \prime }\right )\zeta \left (z^{ \prime }\right ) , \nonumber \end{gather}where the surface mass distribution $\Sigma \left (R^{ \prime }\right )$ can be obtained by interpolating SPARC surface luminosity data and transforming them into surface mass distributions, using appropriate mass-to-light ratios.

For thin disks, the vertical density is described by the delta function $\delta \left (z^{ \prime }_{}\right )$, while for thick disks we used an exponential function $\zeta \left (z^{ \prime }\right ) =\frac{1}{2H_{z}}e^{ -z^{ \prime }/H_{z}}$, where the rescaled parameter $H_{z} =h_{z}/l_{0}$ is connected with the original vertical scale height $h_{z}$. We also adopted the standard relation \cite{Lelli:2016zqa,2010ApJ...716..234B}, $\left (h_{z}/\ensuremath{\operatorname*{kpc}}\right ) =0.196\left (R_{d}/\ensuremath{\operatorname*{kpc}}\right )^{0.633}$, between the scale height $h_{z}$ and the radial scale length $R_{d}$ (available from SPARC data for each galaxy), properly rescaled by using our dimensionless variables.

As mentioned above, the $\Sigma \left (R^{ \prime }\right )$ mass distributions were obtained directly from the SPARC luminosity data. In general, SPARC data include three types of luminosity distributions: a spherically-symmetric \textit{bulge}, a stellar \textit{disk} component, and a \textit{gas} disk component (or simply: bulge, disk, and gas components, for short).

For the disk and gas components, the SPARC surface luminosities $\Sigma _{disk}^{(L)}\left (R^{ \prime }\right )$ and $\Sigma _{gas}^{(L)}\left (R^{ \prime }\right )$, were turned into (rescaled) surface mass distributions, $\Sigma _{disk}\left (R^{ \prime }\right )$ and $\Sigma _{gas}\left (R^{ \prime }\right )$, by using appropriate mass-to-light ratios \cite{McGaugh:2016leg}: $\Upsilon _{disk} \simeq 0.50\ M_{ \odot }/L_{ \odot }$, $\Upsilon _{gas} \simeq 1.33\ M_{ \odot }/L_{ \odot }$ (this value for $\Upsilon _{gas}$ includes also the helium gas contribution).\protect\footnote{
	In more recent papers \cite{Chae:2020omu,Chae:2021dzt}, the gas mass-to-light ratio is taken as $\Upsilon _{gas} =X^{ -1}$, where $X$ is a function of the stellar mass $M_{ \ast }$ in each galaxy, $X =0.75 -38.2\left (\frac{M_{ \ast }}{M_{0}}\right )^{\alpha }$, with $M_{0} =1.5 \times 10^{24}M_{ \odot }$ and $\alpha  =0.22$. The resulting values of $\Upsilon _{gas}$ for the galaxies analyzed in this paper are not much different from the adopted value $\Upsilon _{gas} \simeq 1.33\ M_{ \odot }/L_{ \odot }$, so that we have preferred to use this fixed value of $\Upsilon _{gas}$ in this paper, also to be consistent with our previous analysis in papers II-III.
} We then summed these distributions together, $\Sigma \left (R^{ \prime }\right ) =\Sigma _{gas}\left (R^{ \prime }\right ) +\Sigma _{disk}\left (R^{ \prime }\right )$, and combined this total surface distribution $\Sigma \left (R^{ \prime }\right )$ with the vertical exponential function $\zeta \left (z^{ \prime }\right )$ described above, into the second line of Eq. (\ref{eq8.8}).

For the bulge component, SPARC data are also available in terms of a surface luminosity $\Sigma _{bulge}^{(L)}\left (R^{ \prime }\right )$ which can be turned into a (rescaled) surface mass distribution $\Sigma _{bulge}\left (R^{ \prime }\right )$ by using $\Upsilon _{bulge} \simeq 0.70\ M_{ \odot }/L_{ \odot }$ \cite{Lelli:2016zqa,Lelli:2020pri} and then converted into a spherically-symmetric mass distribution $\rho \left (r^{ \prime }\right )$ by applying Eq. (1.79) in Ref. \cite{2008gady.book.....B}. In this way, for each galaxy studied in this paper, we will have bulge, disk, and gas mass distributions based directly on the experimental SPARC data, instead of using predetermined models, such as exponential, Kuzmin for disk galaxies, and Plummer or other models for spherical components.

The general NFDG potential in the first line of Eq. (\ref{eq8.1}) can then be computed by combining together all the different elements described above in Eqs. (\ref{eq8.2})-(\ref{eq8.8}). In particular, for the case of cylindrical symmetry (thick disk/gas components), we start by integrating analytically the angular part of the expression which includes the Gegenbauer polynomials in Eq. (\ref{eq8.2}) with $\cos \gamma $ replaced by the expression in the third line of Eq. (\ref{eq8.4}), combined with the angular part of the weight in Eq. (\ref{eq8.7}) with the $\alpha $ parameters set as $\alpha _{1} =\alpha _{2} =\frac{D -1}{2}$. This yields the following results:

\begin{gather}c_{l ,D}\left (R^{ \prime } ,z^{ \prime }\right ) ={\displaystyle\int \nolimits_{0}^{2\pi }}\left \vert \sin \phi ^{ \prime }\right \vert ^{\frac{D -3}{2}}\left \vert \cos \phi ^{ \prime }\right \vert ^{\frac{D -3}{2}}C_{l}^{\left (\frac{D}{2} -1\right )}\genfrac{(}{)}{}{}{R^{ \prime }\cos \phi ^{ \prime }}{\sqrt{R^{ \prime 2} +z^{ \prime 2}}}d\phi ^{ \prime } \label{eq8.9} \\
	c_{0 ,D} = -\frac{2^{\frac{5 -D}{2}}\pi ^{3/2}\sec \genfrac{[}{]}{}{}{\pi \left (1 +D\right )}{4}}{\Gamma \genfrac{(}{)}{}{}{5 -D}{4}\Gamma \genfrac{(}{)}{}{}{1 +D}{4}} ;c_{2 ,D} =\frac{2^{\frac{1 -D}{2}}\pi ^{1/2}(D -2)\left [\left (D -2\right )R^{ \prime 2} -2z^{ \prime 2}\right ]\Gamma \genfrac{(}{)}{}{}{D -1}{4}}{\left (R^{ \prime 2} +z^{ \prime 2}\right )\Gamma \genfrac{(}{)}{}{}{1 +D}{4}} ; . . . \nonumber  \\
	c_{1 ,D} =c_{3 ,D} = . . . =0 \nonumber \end{gather}where these functions are identically zero for odd values of $l$, while they can be computed analytically for all even values of $l$ (only the first two are shown in the previous equation).

In order to finish computing the thick-disk potential in the $z =0$ plane, we then combine the previous results in Eq. (\ref{eq8.9}) with the Euler kernel expansion in Eq. (\ref{eq8.2}), supplemented with the transformations in Eq. (\ref{eq8.4}), the weight in Eq. (\ref{eq8.7}) with the choice of parameters $\alpha _{1} =\alpha _{2} =\frac{D -1}{2}$, $\alpha _{3} =1$, and the appropriate mass distribution in the second line of Eq. (\ref{eq8.8}) for each galaxy. Performing the remaining two integrations in $R^{ \prime }$ and $z^{ \prime }$:

\begin{gather}\Phi _{NFDG}^{Cyl .}\left (R\right ) = -\frac{4\sqrt{\pi }\ \Gamma \left (D/2\right )G}{\left (D -2\right )\left [\Gamma \genfrac{(}{)}{}{}{D -1}{4}\right ]^{2}l_{0}}\sum \limits _{l =0}^{\infty }{\displaystyle\int \nolimits_{0}^{\infty }}\zeta \left (z^{ \prime }\right )dz^{ \prime }{\displaystyle\int \nolimits_{0}^{\infty }}c_{l ,D}\left (R^{ \prime } ,z^{ \prime }\right )\Sigma \left (R^{ \prime }\right )\frac{R_{ <}^{l}}{R_{ >}^{l +D -2}}R^{ \prime D -2}dR^{ \prime } \label{eq8.10} \\
	= -\frac{4\sqrt{\pi }\ \Gamma \left (D/2\right )G}{\left (D -2\right )\left [\Gamma \genfrac{(}{)}{}{}{D -1}{4}\right ]^{2}l_{0}}\sum \limits _{l =0}^{\infty }\Bigg\{\left [\frac{1}{R^{l +D -2}}{\displaystyle\int \nolimits_{0}^{R}}\zeta \left (z^{ \prime }\right )dz^{ \prime }{\displaystyle\int \nolimits_{0}^{\sqrt{R^{2} -z^{ \prime 2}}}}c_{l ,D}\left (R^{ \prime } ,z^{ \prime }\right )\Sigma \left (R^{ \prime }\right )\left (\sqrt{R^{ \prime 2} +z^{ \prime 2}}\right )^{l}R^{ \prime D -2}dR^{ \prime }\right ] \nonumber  \\
	+\left [R^{l}{\displaystyle\int \nolimits_{0}^{R}}\zeta \left (z^{ \prime }\right )dz^{ \prime }{\displaystyle\int \nolimits_{\sqrt{R^{2} -z^{ \prime 2}}}^{\infty }}c_{l ,D}\left (R^{ \prime } ,z^{ \prime }\right )\Sigma \left (R^{ \prime }\right )\frac{R^{ \prime ^{D -2}}}{\left (\sqrt{R^{ \prime ^{2}} +z^{ \prime 2}}\right )^{l +D -2}}dR^{ \prime }\right ] \nonumber  \\
	+\left [R^{l}{\displaystyle\int \nolimits_{R}^{\infty }}\zeta \left (z^{ \prime }\right )dz^{ \prime }{\displaystyle\int \nolimits_{0}^{\infty }}c_{l ,D}\left (R^{ \prime } ,z^{ \prime }\right )\Sigma \left (R^{ \prime }\right )\frac{R^{ \prime ^{D -2}}}{\left (\sqrt{R^{ \prime ^{2}} +z^{ \prime 2}}\right )^{l +D -2}}dR^{ \prime }\right ]\Bigg\} . \nonumber \end{gather}

Compared with our previous computation of this thick-disk potential (see Eq. (12) in our paper III), we have corrected a typo in the order of integration for the $R^{ \prime }$ and $z^{ \prime }$ variables, and also improved on the limits of the integrals shown in Eq. (\ref{eq8.10}). In our previous computations \cite{Varieschi:2020dnd,Varieschi:2020hvp}, there was some approximation in the integration limits, which was removed in the current expression shown above.

In particular, to expand the integral in the first line of the above equation, we consider $R_{ <}$ ($R_{ >}$) as the minimum (respectively, maximum) between $R$ and $\sqrt{R^{ \prime ^{2}} +z^{ \prime 2}}$, in view also of Eq. (\ref{eq8.4}). We then split the integral in $z^{ \prime }$ into the two intervals $\left [0 ,R\right ]$ and $\left [R ,\infty \right )$. Considering the first interval for $z^{ \prime } \in \left [0 ,R\right ]$, if $R_{ >} =R$ and $R_{ <} =\sqrt{R^{ \prime 2} +z^{ \prime 2}}$ then $R^{ \prime } \in \left [0 ,\sqrt{R^{2} -z^{ \prime 2}}\right ]$, while if $R_{ >} =\sqrt{R^{ \prime 2} +z^{ \prime 2}}$ and $R_{ <} =R$, then  $R^{ \prime } \in \left [\sqrt{R^{2} -z^{ \prime 2}} ,\infty \right )$ and we obtain the integrals in the second and third lines of Eq. (\ref{eq8.10}). On the other hand, for the second interval $z^{ \prime } \in \left [R ,\infty \right )$, it is always true that $R <\sqrt{R^{ \prime 2} +z^{ \prime 2}}$ for any real value of $R^{ \prime }$, therefore we only need the last integral in the fourth line of Eq. (\ref{eq8.10}).

In view of these corrections of our main formula for the thick-disk NFDG potential, in Appendix \ref{sectiongalacticfive} we will compute again the NFDG results for NGC 7814, NGC 6503, and NGC 3741. In that section, we will show that these revised results are practically equivalent to those previously obtained in our papers II and III. This shows that our previous analysis was not incorrect, but just with some minor approximations in the mathematical formulas, which are now fully corrected in this paper.

In order to compute the gravitational potential $\Phi _{NFDG}^{Sph .}\left (r\right )$ for a spherically-symmetric mass distribution $\rho \left (r\right )$, we can still use our main Eq. (\ref{eq8.1}), but the volume integration must be performed in spherical coordinates. Again, we start by integrating analytically the angular part of the expression which includes the Gegenbauer polynomials in Eq. (\ref{eq8.2}), with $\cos \gamma $ replaced by $\cos \theta ^{ \prime }$, combined with the $\theta ^{ \prime }$ angular part of the weight in Eq. (\ref{eq8.6}) with the $\alpha $ parameters set as $\alpha _{1} =\alpha _{2} =\alpha _{3} =\frac{D}{3}$. This yields the results in the first two lines of the following equation: the $c_{l ,D}$ coefficients are all identically zero for $l =1 ,3 ,5 , . . .$, while they can be computed analytically for $l =0 ,2 ,4 , . . .$ (only the first two are shown in the second line of Eq. (\ref{eq8.11})):
\begin{gather}c_{l ,D} =\int \nolimits_{0}^{\pi }\left \vert \sin \theta ^{ \prime }\right \vert ^{\frac{2D}{3} -1}\left \vert \cos \theta ^{ \prime }\right \vert ^{\frac{D}{3} -1}C_{l}^{\left (\frac{D}{2} -1\right )}\left (\cos \theta ^{ \prime }\right )d\theta ^{ \prime } \label{eq8.11} \\
	c_{0 ,D} =\frac{\pi \csc \genfrac{(}{)}{}{}{\pi D}{6}\Gamma \genfrac{(}{)}{}{}{D}{3}}{\Gamma \left (1 -\frac{D}{6}\right )\Gamma \genfrac{(}{)}{}{}{D}{2}} ;c_{2 ,D} =\frac{\pi \left (\frac{D}{3} -1\right )\csc \genfrac{(}{)}{}{}{\pi D}{6}\Gamma \left (\frac{D}{3}\right )}{\Gamma \left (1 -\frac{D}{6}\right )\Gamma \left (\frac{D}{2} -1\right )} ; . . . \nonumber  \\
	\int \nolimits_{0}^{2\pi }\left \vert \sin \phi ^{ \prime }\right \vert ^{\frac{D}{3} -1}\left \vert \cos \phi ^{ \prime }\right \vert ^{\frac{D}{3} -1}d\phi ^{ \prime } =\frac{2\pi \csc \genfrac{(}{)}{}{}{\pi D}{6}\Gamma \genfrac{(}{)}{}{}{D}{6}}{\Gamma \left (1 -\frac{D}{6}\right )\Gamma \genfrac{(}{)}{}{}{D}{3}} . \nonumber \end{gather}

The integral of the $\phi ^{ \prime }$ angular part of the weight in Eq. (\ref{eq8.6}) with $\alpha _{1} =\alpha _{2} =\frac{D}{3}$ is simply computed and shown in the third line of the last equation. We then combine these results in Eq. (\ref{eq8.11}), with the general integral in Eq. (\ref{eq8.1}), the expansion in Eq. (\ref{eq8.2}), the spherical weight in Eq. (\ref{eq8.6}) with $\alpha _{1} =\alpha _{2} =\alpha _{3} =\frac{D}{3}$, and the spherical mass distribution $\rho \left (r^{ \prime }\right )$ obtained from the bulge surface luminosity $\Sigma _{bulge}^{(L)}\left (R^{ \prime }\right )$ using the procedure described above. The general potential for spherically-symmetric mass distributions can then be written as:

\begin{gather}\Phi _{NFDG}^{Sph .}\left (r\right ) = -\frac{2\pi \Gamma \left (\frac{D}{2} -1\right )G}{\Gamma \left (\frac{D}{3}\right )\Gamma \genfrac{(}{)}{}{}{D}{6}l_{0}}\sum \limits _{l =0 ,2 ,4 , . . .}^{\infty }c_{l ,D}{\displaystyle\int \nolimits_{0}^{\infty }}\rho \left (r^{ \prime }\right )\frac{r_{ <}^{l}}{r_{ >}^{l +D -2}}r^{ \prime D -1}dr^{ \prime } \label{eq8.12} \\
	= -\frac{2\pi \Gamma \left (\frac{D}{2} -1\right )G}{\Gamma \left (\frac{D}{3}\right )\Gamma \genfrac{(}{)}{}{}{D}{6}l_{0}}\sum \limits _{l =0 ,2 ,4 , . . .}^{\infty }c_{l ,D}\left [{\displaystyle\int \nolimits_{0}^{_{r}}}\rho \left (r^{ \prime }\right )\frac{r^{ \prime l}}{r^{l +D -2}}r^{ \prime D -1}dr^{ \prime } +{\displaystyle\int \nolimits_{r}^{\infty }}\rho \left (r^{ \prime }\right )\frac{r^{l}}{r^{ \prime l +D -2}}r^{ \prime D -1}dr^{ \prime }\right ] , \nonumber \end{gather}where the radial integral has been split in two parts in the second line, since $r_{ <}$ ($r_{ >}$) is the minimum (respectively, maximum) of $r$ and $r^{ \prime }$. When $r^{ \prime } \in \left [0 ,r\right ]$ we have $r_{ <} =r^{ \prime }$ and $r_{ >} =r$, while for $r^{ \prime } \in \left [r ,\infty \right )$ we have $r_{ <} =r$ and $r_{ >} =r^{ \prime }$.

If the galaxy being studied does not show a significant spherical bulge component, the computation of the NFDG potential will only include the cylindrical part in Eq. (\ref{eq8.10}), i.e., $\Phi _{NFDG}\left (R\right ) =\Phi _{NFDG}^{Cyl .}\left (R\right )$, but combining the gas and disk distributions together: $\Sigma \left (R^{ \prime }\right ) =\Sigma _{gas}\left (R^{ \prime }\right ) +\Sigma _{disk}\left (R^{ \prime }\right )$. On the contrary, if the spherical bulge is also present, the overall NFDG potential, will combine Eqs. (\ref{eq8.10}) and (\ref{eq8.12}) together:

\begin{equation}\Phi _{NFDG}\left (R\right ) =\Phi _{NFDG}^{Cyl .}\left (R\right ) +\Phi _{NFDG}^{Sph .}\left (R\right ) , \label{eq8.13}
\end{equation}where we can identify the radial coordinates ($r \equiv R$) in the galactic disk plane.

As already remarked in Sect. \ref{sect:REVIEW}, the NFDG gravitational field $\mathbf{g}_{NFDG}\left (R\right )$ can then be computed directly by differentiation of the last equation, as in Eq. (\ref{eq2.8}), and the circular velocities in the galactic disk plane as in Eq. (\ref{eq2.9}). This NFDG\ field $\mathbf{g}_{NFDG}\left (R\right )$ is assumed to correspond to the observed radial acceleration $\mathbf{g}_{obs}\left (R\right )$, which can then be compared with the standard baryonic $\mathbf{g}_{bar}\left (R\right )$, obtained with the same procedure but with a fixed $D =3$ value, or using directly the total baryonic rotational velocity, $V_{bar}\left (R\right )$, available from the SPARC data for each galaxy and with $g_{bar}\left (R\right ) =V_{bar}^{2}\left (R\right )/R$. In Sect. \ref{sectiongalactic}, we used this second option to compute $\mathbf{g}_{bar}$, since it allows for a more direct comparison with the Newtonian behavior.

As in our previous papers, we note that our formulas (\ref{eq8.10}) and (\ref{eq8.12}) for the NFDG gravitational potentials require summing over all non-zero terms for $l =0 ,2 ,4 , . . .$ (terms for odd values of $l$ are identically zero), and we found that these series of functions converge rather quickly over most of the range for $R >0$ (or $r >0$). Following papers II-III, we summed the first six non-zero terms (for $l =0 ,2 ,4 ,6 ,8 ,10$) of our NFDG expansions, although it was probably sufficient to sum only the first three-four terms of each expansion. 

An important improvement of these calculations, compared to papers II-III, was to achieve a good convergence of the series also at low radial distances, thus making our NFDG\ computations reliable over the whole radial range. This was achieved by noting that the NFDG\ gravitational potential is actually independent of $l_{0}$, as it can be seen explicitly by checking Eqs. (\ref{eq8.10}) and (\ref{eq8.12}) where all coordinates are dimensionless (example: $R^{ \prime } \rightarrow R^{ \prime }/l_{0}$, $r^{ \prime } \rightarrow r^{ \prime }/l_{0}$, etc.) and the rescaled mass densities $\Sigma $ and $\rho $ are respectively proportional to $l_{0}^{2}$ and $l_{0}^{3}$. In this way, the NFDG gravitational field computed with Eq. (\ref{eq2.8}) and the circular velocities in Eq. (\ref{eq2.9}) are also independent of the actual choice of the length scale $l_{0}$.

In our previous papers I-III, we used $l_{0} =\sqrt{GM/a_{0}}$ in view of Eq. (\ref{eq2.10}), with $M$ given by the total mass of each galaxy. This choice was usually causing trouble in the convergence of our series at the lowest radial distances. Given that our NFDG main results are actually independent of $l_{0}$ (or that $M$ can be an arbitrary reference mass, as already noted in our previous papers), in this work we opted to reduce the (arbitrary) value of $l_{0}$ by a factor of one thousand, i.e., we adopted:

\begin{equation}l_{0} =10^{ -3}\sqrt{GM/a_{0}} , \label{eq8.14}
\end{equation}
with $M$ being the total mass of the galaxy under study and $a_{0}$ the MOND constant in Eq. (\ref{eq2.11}). This choice of $l_{0}$ fixed the convergence problems at low galactic radial distances and reproduced the same results at higher radial distances obtained in paper III for the first three galaxies studied in NFDG. Therefore, we are confident that our new version of the computation is mathematically sound and that our final results are robust.

\section{NGC 7814, NGC 6503, NGC 3741}
\label{sectiongalacticfive}

In this second appendix, we briefly revisit the cases of NGC 7814, NGC 6503, and NGC 3741, which were already studied in our papers II-III \cite{Varieschi:2020dnd,Varieschi:2020hvp}, following the original RAR results in Ref. \cite{McGaugh:2016leg}. As described in Appendix \ref{sectiongalacticappa}, we have corrected some minor approximations in the limits of the numerical integrations computed with Mathematica software\protect\footnote{
	All numerical computations (and some of the analytical ones) in this paper, were performed with Mathematica, version 13.0, Wolfram Research Inc.
} and improved our numerical analysis at the lowest radial distances. Therefore, we wanted to check these new computations against our previous results for these three galaxies. The main conclusion is that these new results are practically equivalent to those obtained in previous papers II-III, and thus our mathematical analysis is robust.

\begin{figure}\centering 
	\setlength\fboxrule{0in}\setlength\fboxsep{0.1in}\fcolorbox[HTML]{000000}{FFFFFF}{\includegraphics[ width=6.99in, height=8.728805970149253in,]{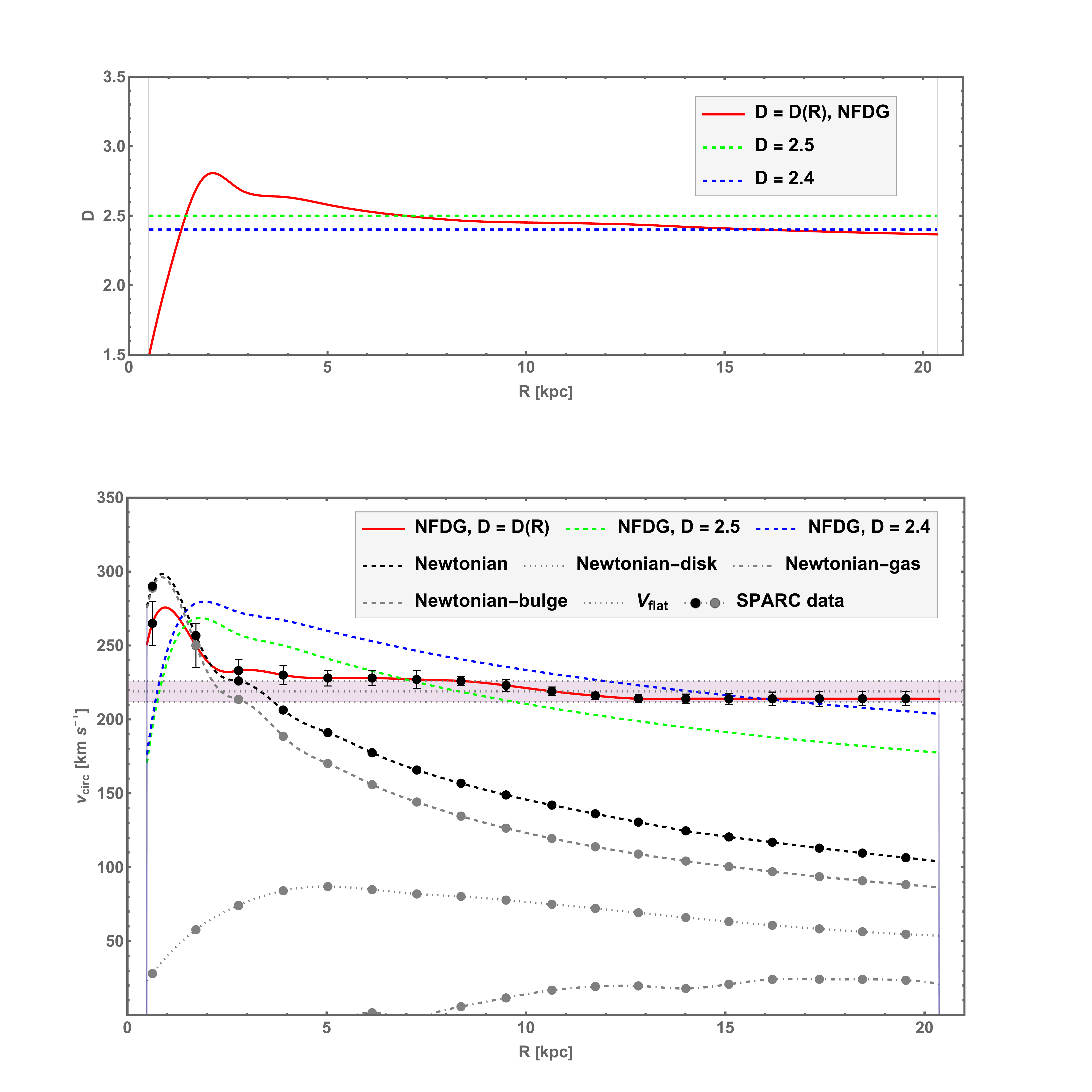}
		}
	\caption{NFDG results for NGC 7814.
		Top panel: NFDG variable dimension $D\left (R\right )$, based directly on SPARC data (red-solid curve), compared with fixed values $D =2.5$ and $D =2.4$ (green and blue-dashed lines). Bottom panel: NFDG rotation curves (circular velocity vs. radial distance) compared to the original SPARC data (black circles with error bars). The NFDG best fit for the variable dimension $D\left (R\right )$ is shown by the red-solid line, while NFDG fits for fixed values $D =2.5$ and $D =2.4$ are shown respectively by the green and blue dashed lines. Also shown: Newtonian rotation curves (different components - gray lines, total - black dashed line) with corresponding SPARC data (gray/black circles), and asymptotic flat velocity band (horizontal gray band).}
		\label{figure:NGC7814_1}
	\end{figure}

NGC 7814 is a spiral galaxy in the constellation Pegasus, which has a dominating bulge component and less prominent disk and gas components. SPARC data for this galaxy include the following:\ distance $D =\left (14.40 \pm 0.66\right )$ \textrm{Mpc}, disk scale length $R_{d} =2.54\ kpc =7.84 \times 10^{19}\mbox{m}$, asymptotically flat rotation velocity $V_{f} =\left (218.9 \pm 7.0\right )\mbox{km}/\mbox{s}$, and computed galactic masses $M_{bulge} =5.76 \times 10^{40}\mbox{kg}$, $M_{disk} =2.41 \times 10^{40}\mbox{kg}$, $M_{gas} =2.80 \times 10^{39}\mbox{kg}$, $M_{total} =8.45 \times 10^{40}\mbox{kg}$. The NFDG results for this galaxy are illustrated in Fig. \ref{figure:NGC7814_1}, with radial limits set at $R_{\min } =0.492\ \ kpc
$ and $R_{\max } =20.4\ \ kpc$.

The results are presented in the usual way: the NFDG variable dimension $D\left (R\right )$ is shown in the top panel as a red-solid curve, with the corresponding NFDG fit to the SPARC experimental data shown by the red-solid curve in the bottom panel. In the same figure, we also show the NFDG circular velocity fits for $D =2.5$ (green-dashed curves) and $D =2.4$ (blue-dashed curves), with the $D =2.4$ curve now being a close fit at the largest distances.

This galaxy compares well with the previous cases for NGC 5033 and NGC 6674, the other two galaxies with a strong bulge component, with the dimension progressively decreasing at larger distances toward an almost constant value at the  largest radii. The only difference for NGC 7814 is the peculiar behavior of $D\left (R\right )$ at lowest radial distances - decreasing toward $D \approx 1.5$ - which might be an unphysical outcome of our numerical analysis. Comparing these new results for NGC 7814 with those obtained in paper III\ using our previous version of the numerical computations (see Fig. 1 in Ref. \cite{Varieschi:2020hvp}), we do not notice any significant difference in the main NFDG curves, thus confirming that our results are not affected by the slight changes in our numerical computations discussed in Appendix \ref{sectiongalacticappa}.

\begin{figure}\centering 
	\setlength\fboxrule{0in}\setlength\fboxsep{0.1in}\fcolorbox[HTML]{000000}{FFFFFF}{\includegraphics[ width=6.99in, height=8.728805970149253in,]{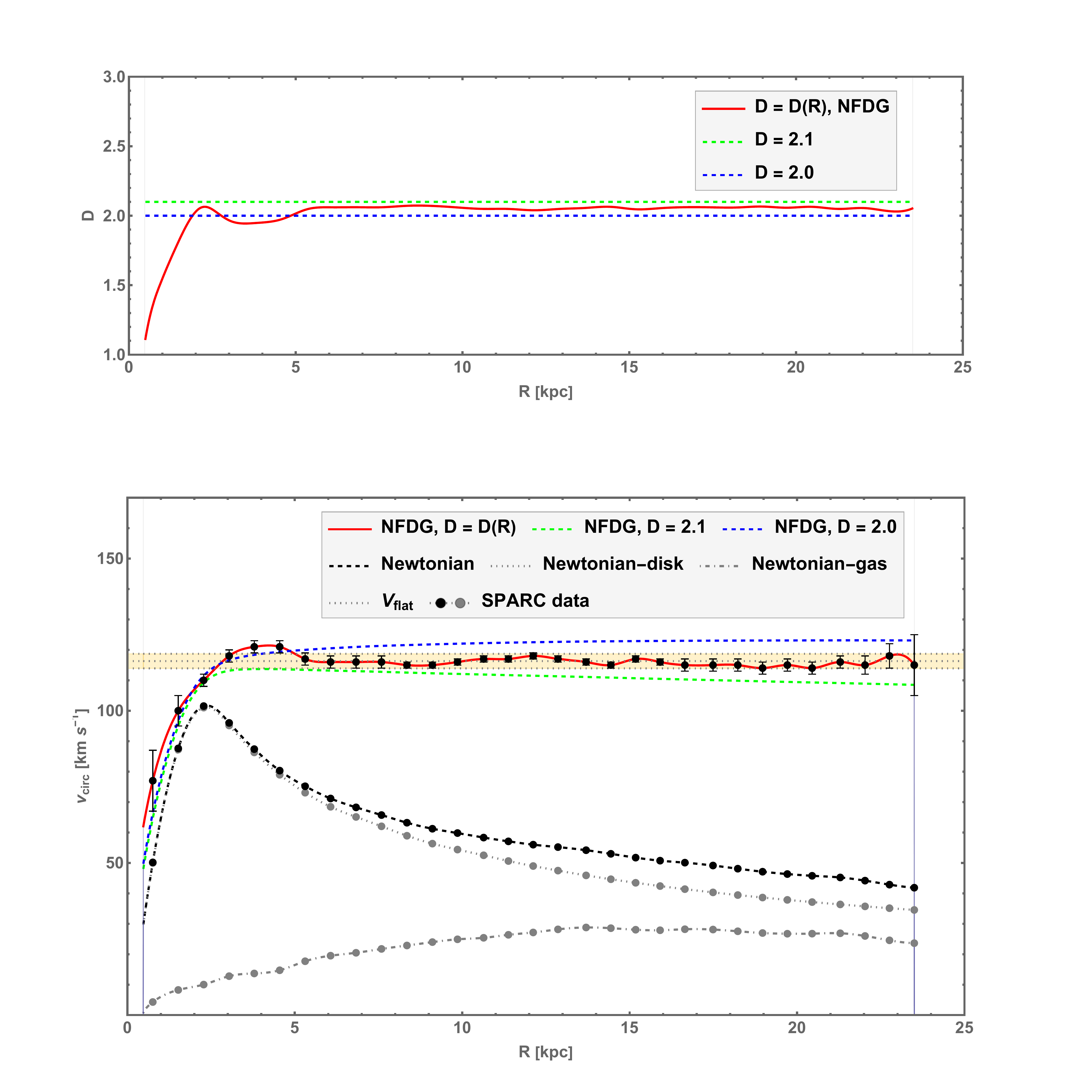}
	}
	\caption{NFDG results for NGC 6503.
		Top panel: NFDG variable dimension $D\left (R\right )$, based directly on SPARC data (red-solid curve), compared with fixed values $D =2.1$ and $D =2.0$ (green and blue-dashed lines). Bottom panel: NFDG rotation curves (circular velocity vs. radial distance) compared to the original SPARC data (black circles with error bars). The NFDG best fit for the variable dimension $D\left (R\right )$ is shown by the red-solid line, while NFDG fits for fixed values $D =2.1$ and $D =2.0$ are shown respectively by the green and blue dashed lines. Also shown: Newtonian rotation curves (different components - gray lines, total - black dashed line) with corresponding SPARC data (gray/black circles), and asymptotic flat velocity band (horizontal gray band).}
	\label{figure:NGC6503_1}
\end{figure}

NGC 6503 is a field dwarf spiral galaxy in the constellation Draco, which has a dominating stellar disk component, a less prominent gas component, and no bulge. As such, it is comparable with NGC 5055 and NGC 1019 analyzed above, in Sects. \ref{sectiongalacticthree} and \ref{sectiongalacticfour}. SPARC data for this galaxy include the following:\ distance $D =\left (6.26 \pm 0.31\right )$ \textrm{Mpc}, disk scale length $R_{d} =2.16\ kpc =6.67 \times 10^{19}\mbox{m}$, asymptotically flat rotation velocity $V_{f} =\left (116.3 \pm 2.4\right )\mbox{km}/\mbox{s}$, and integrated galactic masses $M_{disk} =1.26 \times 10^{40}\mbox{kg}$, $M_{gas} =4.60 \times 10^{39}\mbox{kg}$, $M_{total} =1.72 \times 10^{40}\mbox{kg}$. The NFDG results for this galaxy are illustrated in Fig. \ref{figure:NGC6503_1}, with radial limits set at $R_{\min } =0.476\ \ kpc
$ and $R_{\max } =23.5\ \ kpc$.

Comparing this galaxy with the previous cases of NGC 5055 and NGC 1090, the other two disk-dominated galaxies without a bulge component, we notice that the variable dimension for NGC 6503 remains in the range between $D \simeq 2.0$ and $D \simeq 2.1$ for most radial distances. At the lowest radii, we observe $D \approx 1.0 -1.5$, and comparing these new results for NGC 6503 with those obtained in paper III\  with our previous version of the numerical computations (see Fig. 3 in Ref. \cite{Varieschi:2020hvp}), we observe an improvement in our main fit at the lowest distances, due to the better numerical computation described in Appendix \ref{sectiongalacticappa}.

\begin{figure}\centering 
	\setlength\fboxrule{0in}\setlength\fboxsep{0.1in}\fcolorbox[HTML]{000000}{FFFFFF}{\includegraphics[ width=6.99in, height=8.728805970149253in,]{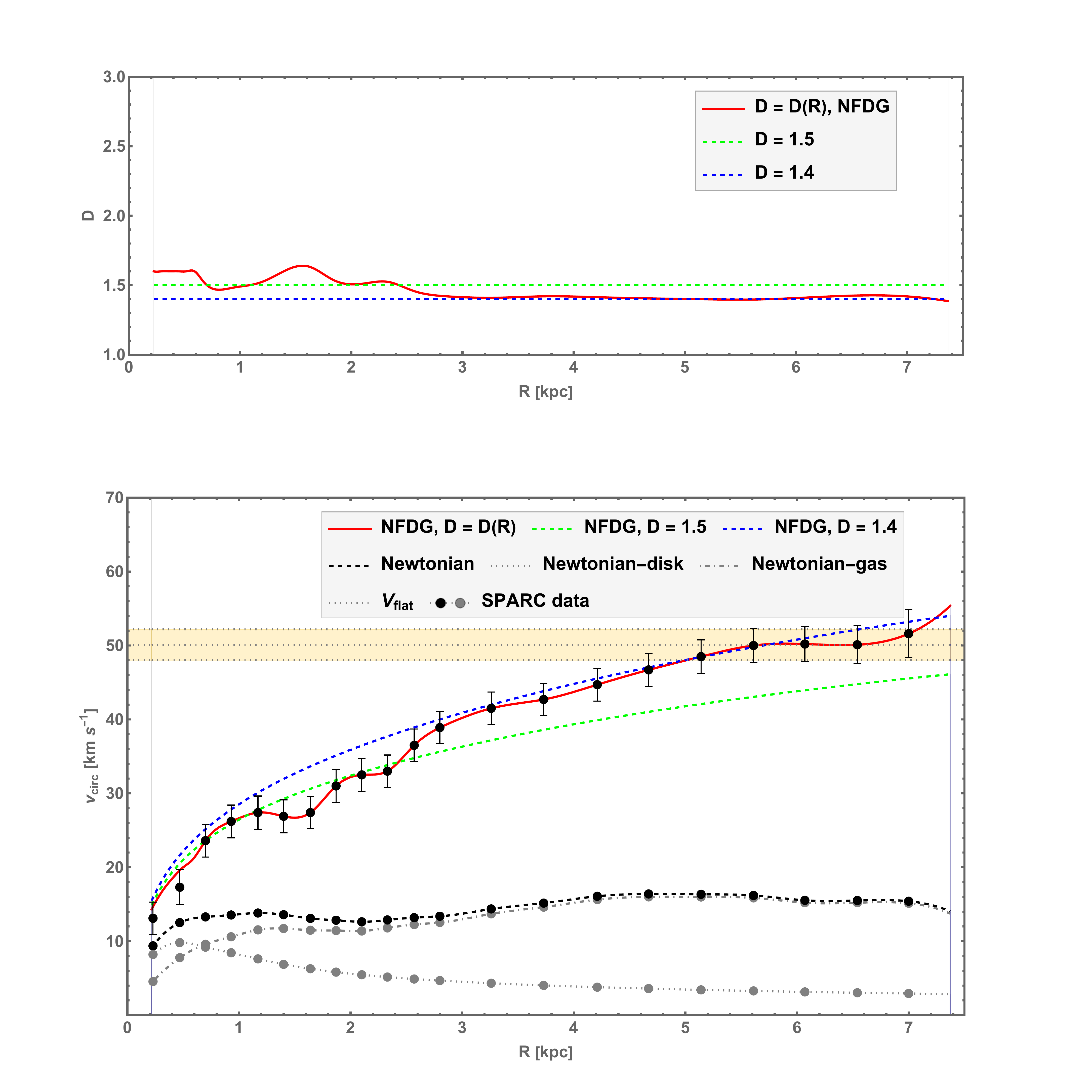}
	}
	\caption{NFDG results for NGC 3741.
		Top panel: NFDG variable dimension $D\left (R\right )$, based directly on SPARC data (red-solid curve), compared with fixed values $D =1.5$ and $D =1.4$ (green and blue-dashed lines). Bottom panel: NFDG rotation curves (circular velocity vs. radial distance) compared to the original SPARC data (black circles with error bars). The NFDG best fit for the variable dimension $D\left (R\right )$ is shown by the red-solid line, while NFDG fits for fixed values $D =1.5$ and $D =1.4$ are shown respectively by the green and blue dashed lines. Also shown: Newtonian rotation curves (different components - gray lines, total - black dashed line) with corresponding SPARC data (gray/black circles), and asymptotic flat velocity band (horizontal gray band).}
	\label{figure:NGC3741_1}
\end{figure}

Finally, NGC 3741 is an irregular galaxy in the constellation Ursa Major, which has a dominating gas component, a less prominent disk component, and no bulge. Therefore, it is different from all the other galaxies analyzed with NFDG techniques, and this might be the reason for the different type of dimension function $D\left (R\right )$ obtained in this case. SPARC data for this galaxy include the following:\ distance $D =\left (3.21 \pm 0.17\right )$ \textrm{Mpc}, disk scale length $R_{d} =0.20\ kpc =6.17 \times 10^{18}\mbox{m}$, asymptotically flat rotation velocity $V_{f} =\left (50.1 \pm 2.1\right )\mbox{km}/\mbox{s}$, and galactic masses $M_{disk} =2.68 \times 10^{37}\mbox{kg}$, $M_{gas} =4.80 \times 10^{38}\mbox{kg}$, $M_{total} =5.06 \times 10^{38}\mbox{kg}$. The NFDG results for this galaxy are illustrated in Fig. \ref{figure:NGC3741_1}, with radial limits set at $R_{\min } =0.218\ \ kpc
$ and $R_{\max } =7.37\ \ kpc$.

As remarked above, the overall NFDG results for this galaxy are quite different from the previous cases, probably due to the strong gas component of this galaxy. We observe $D \approx 1.4 -1.5$ for most of the radial range, so that we might conclude that a strong galactic gas component effectively reduces the fractional dimension of the galaxy well below $D \approx 2$. Further analysis of other gas dominated galaxies will be needed to confirm this hypothesis. Comparing these new results for NGC 3741 with those obtained in paper III\  with our previous version of the numerical computation (see Fig. 5 in Ref. \cite{Varieschi:2020hvp}), we only notice an improvement in our main fit a the lowest distance, but no other significant changes.

In summary, repeating the NFDG analysis with our improved numerical simulations for these last three galaxies has resulted in notable improvements at the lowest radial distance, but no other practical differences at larger distances. Our past analysis remains correct and the improvements introduced here will be used in future studies of other galaxies.

\section{NGC 5055 - extended analysis}
\label{sectiongalacticdeep5055}

In this final appendix, we extend our analysis of NGC 5055 in order to address certain issues related to our general NFDG methods. We limit this additional analysis just to NGC 5055 for brevity, but we checked that similar results are also obtained for the other galaxies considered in this paper.

As mentioned in Sect. \ref{sect:intro}, in addition to our standard NFDG analysis, we can also use for the first time a direct determination of the variable dimension function based on the concept of mass dimension of a fractal system and on the individual galactic mass distributions.

For an isotropic fractal material, the mass dimension $D$ is usually defined \cite{bookTarasov,Tarasov:2014fda,TARASOV2015360} as: $M_{D}\left( W_{B} \right)=M_{0}{\left( \frac{R}{R_{0}} \right)}^D$, 
where $M_{D}$ is the mass of a ball region $W_{B}$ of a fractal medium of radius $R$, $R_{0}$ is a characteristic scale of the fractal medium and $M_{0}$ is the mass of a ball of radius $R_{0}$. For $D=3$ we recover the usual result that the mass of a ball of uniform density scales like the cube of the radius $R$.

Since our galactic fractal structures are more disk-shaped than spherically-shaped and their mass distributions are given in terms of a surface distribution, $\Sigma_{tot} \left (R^{ \prime }\right ) =\Sigma _{gas}\left (R^{ \prime }\right ) +\Sigma _{disk}\left (R^{ \prime }\right )+\Sigma _{bulge}\left (R^{ \prime }\right )$, for the three main components (see Appendix \ref{sectiongalacticappa} for details), we can modify the definition of the mass dimension (denoted by $D_{m}$ in the following) as follows:
\begin{equation}M\left( R \right) \approx M_{0}{\left( \frac{R}{R_{0}} \right)}^{D_{m}(R)-1} , \label{eq7.1}
\end{equation}
since for $D_{m}=3$ the mass of a surface distribution should scale as $R^2$.

In the previous equation, we now consider $D_{m}=D_{m}(R)$ as typical in NFDG and the masses $M(R)$ and $M_{0}$ can be computed directly from the total surface distribution: $M\left( R \right)=2\pi\int_{0}^{R} \Sigma_{tot}\left( R' \right)R'dR'$ and $M_{0}\equiv M\left( R_{0} \right)=2\pi\int_{0}^{R_0} \Sigma_{tot}\left( R' \right)R'dR'$. Eq. \ref{eq7.1} can then be solved directly for the variable mass dimension:
\begin{equation}D_{m}\left( R \right) \approx 1+\ln\left[ \frac{M\left( R \right)}{M\left( R_0 \right)} \right]/\ln\left[ \frac{R}{R_0} \right] , \label{eq7.2}
\end{equation}
and will provide a prediction of the galactic variable dimension which is, in principle, independent of the observed galactic circular velocities.

However, the shape of the function $D_{m}(R)$ still depends on the choice of the scale length $R_{0}$ (and related reference mass $M_{0}$) which can be considered as a free parameter, or can be set by matching Eq. \ref{eq7.1} with at least one observed value of the galaxy circular velocity. Therefore, we set:
\begin{equation}\frac{M_{D}\left( R_{data} \right)}{M_{D}\left( R_{0} \right)}=\left( \frac{R_{data}}{R_{0}} \right)^{D\left( R_{data} \right)-1} , \label{eq7.3}
\end{equation}
where $R_{data}$ is chosen as the radial distance of one of the SPARC data points in the flat velocity region of the galaxy being considered, $D\left( R_{data} \right)$ is computed with our standard NFDG methods and the equation is solved numerically for the scale radius $R_{0}$.

\begin{figure}\centering 
	\setlength\fboxrule{0in}\setlength\fboxsep{0.1in}\fcolorbox[HTML]{000000}{FFFFFF}{\includegraphics[width=6.99in, height=8.728805970149253in,]{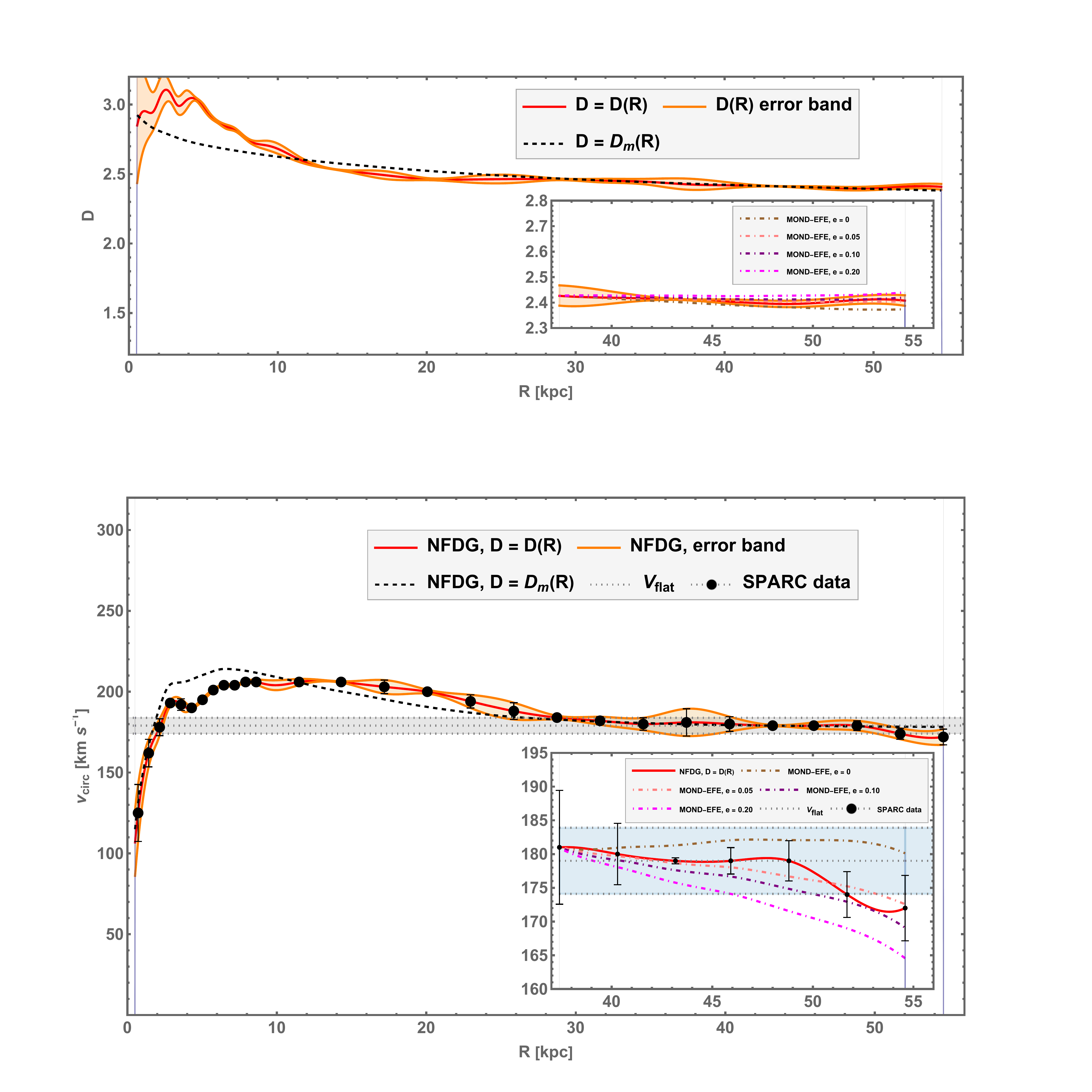}
	}
	\caption{Extended NFDG results for NGC 5055.
		Top panel: NFDG variable dimension $D\left (R\right )$, based directly on SPARC data (red-solid curve) and related error band (orange-shaded area), compared with variable mass dimension $D_{m}(R)$ (black-dashed line). Bottom panel: NFDG rotation curves (circular velocity vs. radial distance) compared to the original SPARC data (black circles with error bars). The NFDG best fit for the variable dimension $D\left (R\right )$ is shown by the red-solid line, while an alternative NFDG fit based on the variable mass dimension $D_{m}(R)$ is shown by the black-dashed line. Also shown in the insets in both panels: MOND-EFE analysis of the outer radial region, for different values of the EFE parameter $e$.}
	\label{figure:NGC5055_extra}
\end{figure}

Figure \ref{figure:NGC5055_extra} shows the results of this extended analysis for NGC 5055 and improves on our previous Fig. \ref{figure:NGC5055_1}. In addition to our standard NFDG fit and related dimension function $D(R)$ (red-solid curves, in both panels), we also show in the top panel the variable mass dimension $D_{m}(R)$ obtained using Eqs. (\ref{eq7.1})-(\ref{eq7.3}) and the related rotation velocity fit in the bottom panel (black-dashed curves).

Following Eq. (\ref{eq7.3}), we set the free parameter $R_{0}$ to match just one of the SPARC data (the fifth to last point, which best matches the SPARC flat velocity for this galaxy). Apart from this choice, the results just depend on the surface mass distributions. Although the resulting rotation velocity fit (black-dashed curve) is not as perfect as our standard NFDG fit (red-solid curve), it is nevertheless a rather good fit over most of the radial range. We will improve this methodology in future studies, in order to fully address the issue of the falsifiability of NFDG.

In the same figure, we also include a possible error band for our main NFDG dimension function $D(R)$. This was done by considering the error band in the bottom panel (orange-shaded area), which is based on the published error bars of the SPARC data shown in the same panel of the figure. Using our NFDG simulations, this circular velocity error band is then transformed into the related error band for $D(R)$ in the top panel of the figure (orange-shaded area), which can give a measure of the uncertainty of the dimension function. As it can be seen from the top panel of Fig. \ref{figure:NGC5055_extra}, this uncertainty is of the order $\Delta D \lesssim 0.1$ over most of the radial range and we expect this to be a typical value also for the other galaxies studied in this paper.

Finally, in the inset figures in both panels of Fig. \ref{figure:NGC5055_extra} we compare our NFDG results with the MOND-EFE computations following Eq. (\ref{eq3.2}). In the inset of the bottom panel, we consider the outer radial region of the last seven SPARC data points and compute the MOND-EFE curves for $e=0$ (non-EFE case) and $e=0.05, 0.10, 0.20$ (EFE-cases). These results from Eq. (\ref{eq3.2}) were in part normalized, in order to match the seventh to last data point, and are also compared with our main NFDG fit (red-solid curve) in the same inset.

As expected, the MOND-EFE curves for $e=0.05-0.10$ can fit the data better than the $e=0$ non-EFE curve (Chae at al. \cite{Chae:2020omu} report $e=0.054$ for this galaxy), while the NFDG red-solid curve does not include any EFE correction terms, since our model does not admit the EFE. However, in the top-panel inset we computed also the equivalent NFDG dimension functions corresponding to the four MOND-EFE curves in the bottom-panel inset.

This was done to show how sensitive our $D(R)$ computation is, with respect to the different MOND-EFE curves. As seen in this top-panel inset, the differences in the dimension function over this outer radial range are minimal. It is true that the systematic downward deviation of the MOND-EFE rotation velocity curves in the bottom-panel inset corresponds to an opposite upward deviation of the related dimension functions in the top-panel inset, but these deviations of the dimension function are minimal and comparable to the estimated error band.

In other words, a possible MOND-EFE effect described in terms of the NFDG variable dimension $D(R)$ would only yield minor changes in the slope of this function in the outer radial regions and not a major change of the actual value of the variable dimension at larger radii, which might easily allow to distinguish between EFE/non-EFE cases within the NFDG paradigm. This seems to confirm our conclusions about NFDG and the EFE, as outlined in the main sections of this paper.

\bibliographystyle{apsrev4-2}
\bibliography{RFDGmainNotes}
\end{document}